\documentclass[usenatbib]{mnras}

\usepackage[T1]{fontenc}

\usepackage{amssymb}
\usepackage[normalem]{ulem}
\usepackage{amsmath}
\usepackage{relsize}
\usepackage[pdftex]{graphicx}
\usepackage{mathptmx}
\usepackage{color}
\usepackage{xcolor}

\usepackage{enumitem}

\newcommand{\be}{\begin{equation}}
\newcommand{\ee}{\end{equation}}

\makeatletter
\newcommand*{\rom}[1]{\expandafter\@slowromancap\romannumeral #1@}
\makeatother

\newcommand{\cosmosis}{\textsc{cosmosis}\,}
\newcommand{\acamb}{\textsc{axionCAMB}\,}
\newcommand{\emcee}{\textsc{emcee}\,}
\newcommand{\HMC}{\textsc{HMCode}\,}

\renewcommand{\vec}[1]{{\mathbf{#1}}}

\begin{document}

\label{firstpage} % <===================================================
\pagerange{\pageref{firstpage}--\pageref{lastpage}}

\title{Fuzzy Dark Matter and the Dark Energy Survey Year 1 Data}
\author[M. Dentler, D. J. E. Marsh, R. Hlo{\v z}ek,  A. Lagu\"{e}, K. K. Rogers, and D. Grin] {Mona Dentler$^1$, David~J.~E.~Marsh$^{2}$\thanks{Corresponding Author: david.j.marsh@kcl.ac.uk}, Ren\'{e}e Hlo{\v z}ek$^{3,4}$, \newauthor Alex Lagu\"{e}$^{3,4,5}$, Keir K. Rogers$^3$, Daniel Grin$^{6}$.
\\$^1$Institut f\"ur Astrophysik, Friedrich-Hund-Platz 1, 37077 G\"ottingen, Germany
\\$^2$King's College London, Strand, London, WC2R 2LS, United Kingdom
\\$^3$Dunlap Institute for Astronomy and Astrophysics, University of Toronto, 50 St. George Street, Toronto, Ontario, M5S 3H4, Canada
\\$^4$Department of Astronomy and Astrophysics, University of Toronto, 50 St. George Street, Toronto, Ontario, M5S 3H4, Canada
\\$^5$Canadian Institute for Theoretical Astrophysics, University of Toronto, 60 St. George St., Toronto, ON, M5S 3H8, Canada\\
$^6$Department of Physics and Astronomy, Haverford College, 370 Lancaster Avenue, Haverford, PA 19041, United States}

\date{\today}

\maketitle

\begin{abstract}

Gravitational weak lensing by dark matter halos leads to a measurable imprint in the shear correlation function of galaxies. Fuzzy dark matter (FDM), composed of ultralight axion-like particles of mass $m\sim 10^{-22}\text{ eV}$, suppresses the matter power spectrum and shear correlation with respect to standard cold dark matter. We model the effect of FDM on cosmic shear using the optimised halo model \textsc{HMCode}, accounting for additional suppression of the mass function and halo concentration in FDM as observed in $N$-body simulations. We combine Dark Energy Survey year 1 (DES-Y1) data with the \emph{Planck} cosmic microwave background anisotropies to search for shear correlation suppression caused by FDM. We find no evidence of suppression compared to the preferred cold DM model, and thus set a new lower limit to the FDM particle mass. Using a log-flat prior and marginalising over uncertainties related to the non-linear model of FDM, we find a new, independent 95\% C.L. lower limit $\log_{10}m>-23$ combining \emph{Planck} and DES-Y1 shear, an improvement of almost two orders of magnitude on the mass bound relative to CMB-only constraints. Our analysis is largely independent of baryonic modelling, and of previous limits to FDM covering this mass range. Our analysis highlights the most important aspects of the FDM non-linear model for future investigation. The limit to FDM from weak lensing could be improved by up to three orders of magnitude with $\mathcal{O}(0.1)$ arcmin cosmic shear angular resolution, if FDM and baryonic feedback can be simultaneously modelled to high precision in the halo model. 
\end{abstract}

%
%
%
%%%%%%%%%%%%%%%%%%%%%%%%%%%%%%%%%%%%%%%%%%%%%%%%%%%%%%%%%%%%%%%%%%%%%%%%%%%%%%%%%
%%%%%%%%%%%%%%%%%%%%%%%%%%%%%%%%%%%%%%%%%%%%%%%%%%%%%%%%%%%%%%%%%%%%%%%%%%%%%%%%%
\section{Introduction}
\label{sec:intro}

Dark matter (DM) is one of the most pressing issues in modern particle physics and cosmology, with its existence confirmed by a range of observations covering scales from the galactic neighbourhood to the cosmos~\citep{Planck:2018vyg,deSalas:2020hbh}. At one extreme end of DM candidates reside primordial black holes (PBHs), with macroscopic masses measured in solar masses, $M_\odot$. Observations constrain the parameter space of PBHs tightly, leaving a single window around $M\approx 10^{-12}M_\odot$ where they can compose all of the DM with a monochromatic mass function, and with tight constraints on the PBH fraction across the rest of the parameter space~\citep{Green:2020jor}. At the other extreme end of DM parameter space are ultralight bosons, including pseudoscalars such as the axion~\citep{Peccei:1977hh,Weinberg:1977ma,Wilczek:1977pj,Abbott:1982af,Preskill:1982cy,Dine:1982ah}, scalars~\citep[e.g.][]{Turner:1983he,Li:2013nal}, vectors~\citep[e.g.][]{Graham:2015rva}, and tensors~\citep[e.g][]{Babichev:2016bxi}. 

In this work we are concerned with ultralight scalars and pseudoscalars that can be treated as having vanishing non-gravitational self interactions at early times \citep{Marsh:2021lqg}. We consider DM composed of a scalar field, $\phi$, with potential $V(\phi)=m^2\phi^2/2$, where $m$ is the particle mass, and neglect all other DM interactions (hence the model applies to scalars and pseudoscalars equally). We consider masses in the range of $10^{-27}\text{ eV}\lesssim m\lesssim10^{-19}\text{eV}$. We  refer to such particles as Fuzzy Dark Matter (FDM), or ``ultralight bosonic dark matter'' (UBDM). Ultralight axions (ULAs) provide one particle physics model for FDM, valid in the mass range of interest for ``decay constants'' in the range $f_a\approx 10^{17}\text{ GeV}$. Such ULAs are a feature of the string theory landscape~\citep{Arvanitaki:2009fg,Marsh:2015xka,Hui:2016ltb,Mehta:2021pwf,Cicoli:2021gss}, and can also be realised in field theory models~\citep{Kim:2015yna,Davoudiasl:2017jke}.

In the context of cosmology, the key phenomenological  feature of FDM/ULAs is the existence of a scale depending on the particle mass $m$, called the Jeans scale $k_J(m)$~\citep{Khlopov:1985jw}. At smaller wavenumbers $k<k_J$, FDM is phenomenologically virtually indistinguishable  from cold DM (CDM). However, at higher wavenumbers $k>k_J$, the scalar field gradient energy leads to an effect called ``quantum pressure'',~\footnote{Erroneously, since it is neither quantum, nor a pressure.} which is absent in the case of CDM. The presence of quantum pressure below the Jeans scale counteracts gravity and results in a suppression of the matter density power spectrum, $P(k)$~\citep[][]{Hu:2000ke,Marsh:2010wq,Marsh:2015xka,Hui:2016ltb,Marsh:2021lqg}. 

Our evidence for DM comes entirely from its gravitational interactions. So too come the most widely applicable and generic constraints on the properties of DM, in particular the lower bound to the particle mass, $m$. Cosmic microwave background (CMB) temperature, polarisation, and gravitational lensing anisotropies measured by the \emph{Planck} satellite~\citep{Planck:2019nip} establish the lower limit $m\gtrsim \mathcal{O}(\text{few})\times 10^{-25}\text{ eV}$ for the dominant component of DM~\citep{Hlozek:2014lca,Hlozek:2017zzf,Poulin:2018dzj,LinaresCedeno:2020dte}. The CMB bound is extremely robust, since it relies only on linear physics of cosmological perturbation theory and decoupling of the photon-baryon plasma, and marginalises over uncertainties in the cosmological parameters (baryon density, DM density, primordial power, Hubble parameter, and CMB optical depth). 

Extending the lower limit to the FDM mass requires probes of the power spectrum on scales smaller than those accessible from the CMB, moving towards the quasi-linear and fully non-linear regimes of structure formation. Many independent probes have been considered, including from the epoch of reionisation \citep[e.g.][]{Bozek:2014uqa,Sarkar:2015dib,Schive:2015kza,Corasaniti:2016epp}. In these studies, cosmological and systematic uncertainties are left largely unaccounted for. A more rigorous approach to small and non-linear scales is required.

One such rigorous approach has been developed for the Lyman-$\alpha$ forest flux power spectrum~\citep[see e.g.][]{Narayanan:2000tp,McDonald:2004eu,Viel:2013fqw,Irsic:2017ixq}. The highest-resolution Lyman-$\alpha$ forest probes the matter power spectrum as traced by the intergalactic medium (IGM) for line-of-sight velocity wavenumbers $k_\mathrm{f} < 0.2\, \text{s}\,\text{km}^{-1}$ and redshifts $4.2<z<5$ \citep{2019ApJ...872..101B}. The most accurate model on such small scales is a cosmological hydrodynamical simulation \citep[e.g.][]{2015MNRAS.446.3697L}, ideally run in pairs in order to suppress sample variance \citep{Anderson:2018zkm}.

The computational cost of such simulations necessitates, for parameter inference, the use of \emph{emulators}: interpolation methods in high-dimensional parameter spaces\footnote{This is despite the computational efficiency in Lyman-$\alpha$ forest simulations where hot and dense gas particles are converted into collisionless particles. This has negligible impact on Lyman-$\alpha$ forest statistics, which are sourced in the lower-density IGM \citep{10.1111/j.1365-2966.2004.08224.x}.} \citep[e.g.][]{2010ApJ...713.1322L}. In particular, \cite{Bird:2018efe,2019JCAP...02..031R,Rogers:2020ltq,Rogers:2020cup} built Lyman-$\alpha$ forest emulators capable of marginalising the astrophysical uncertainties of IGM temperature, energy deposition in the IGM, and the strength of the ionising background, in addition to cosmological parameters, and the FDM mass, leading to the rigorous bound $m>2\times 10^{-20}\text{ eV}$ at 95\% credibility. The methodology to marginalise uncertainties improves on previous Lyman-$\alpha$ forest bounds on ULAs~\citep{Irsic:2017yje,Armengaud:2017nkf,Kobayashi:2017jcf}. 

A shortcoming of the constraints to FDM from the Lyman-$\alpha$ forest is that the full power of a joint likelihood analysis with the CMB has not yet been exploited \citep[although see e.g.][for the case of other cosmological parameters]{Seljak:2006bg}. Studies using the Lyman-$\alpha$ forest neglect the interplay of the quantum pressure and baryons in ways which may weaken bounds \citep{Zhang2018:TheImportance}. However, large-scale structure simulations including FDM dynamics indicate that, for current data, the effect is only on the order of a few percent \citep{Nori:2018pka}. Further, existing Lyman-$\alpha$ forest bounds neglect the large-scale fluctuations (\(\sim 40\,\mathrm{Mpc}\)) in IGM temperature and ionisation arising from a spatially-inhomogeneous reionisation \citep{Hui:2016ltb}. Although, initial studies \citep{2019MNRAS.490.3177W, Molaro:2021tdz} suggest that the effect on the small scales (sub-Mpc) that drive dark matter bounds are currently statistically negligible. Nonetheless, this motivates complementary and rigorous probes of the dark matter nature, in particular exploiting scales intermediate to those traced by the CMB primary anisotropies (\(\sim \mathrm{Gpc}\)) and the high-resolution Lyman-$\alpha$ forest (\(\sim \mathrm{Mpc}\)).

%%%%%%%%%%%%%%%%%%%%%%%%%%%
\begin{figure}
 \begin{center}
 \includegraphics[width=.45\textwidth]{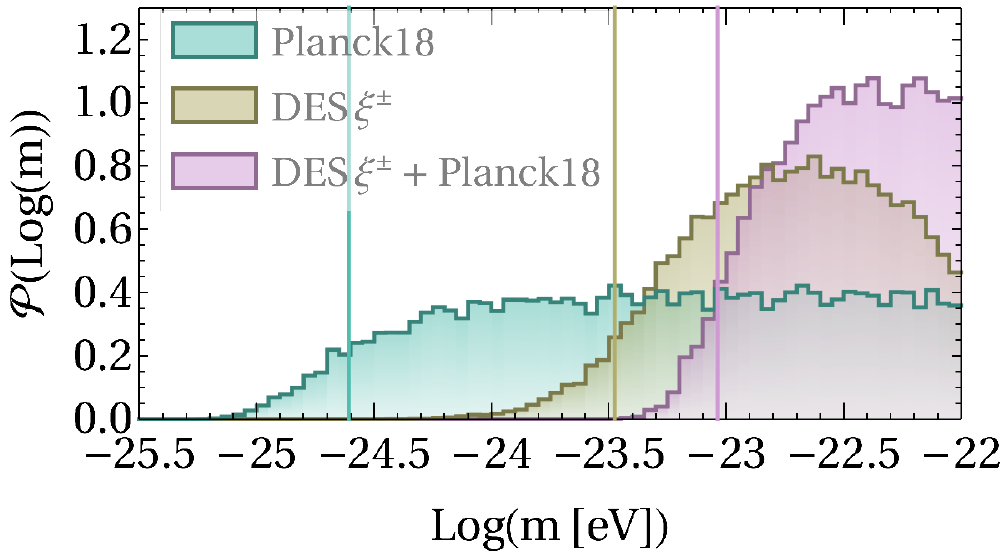}
 \caption{One dimensional posterior on the FDM particle mass, $m$, assuming a log-flat prior. The DES cosmic shear correlation function, $\xi_\pm$ probes scales one order of magnitude smaller than those probed by the \emph{Planck} CMB anisotropies. Thus, given the absence of evidence for scale dependence of $\xi_\pm$ induced by the FDM Jeans scale, $k_{J,eq}\propto m^{1/2}$, Eq.~\eqref{eqn:kj_eq}, we are able to place a new lower limit to $m$ that is almost two orders of magnitude tighter than the limit from the CMB alone. This new limit is unaffected by a variety of uncertainties associated to the non-linear model. The small peak in the posterior using DES alone arises due to degeracies that are broken by the combination with CMB data, and is not statistically significant.}\label{fig:posterior_1d}
\end{center}
\end{figure}
%%%%%%%%%%%%%%%%%%%%%%%%%
An alternative and complementary observable that probes small and non-linear scales is provided by the galaxy shear weak lensing power spectrum, and is the subject of the present work. In particular, we make use of the Dark Energy Survey Year 1~\citep[DES-Y1,][]{Abbott:2017wau} weak lensing likelihood implemented in the publicly available \cosmosis~\citep{cosmosis} analysis framework.~\footnote{While this work was in preparation the DES-Y3 results were released as preprints~\citep{DES:2021zxv,DES:2021bpo}. We will study these in a future publication, once the DES-Y3 likelihoods have been publicly released.} A virtue of weak lensing is that the theoretical input is the total matter power spectrum, $P(k)$, offering a direct probe of the clustering not just of a tracer such as hydrogen gas in the IGM, but of the DM itself. 

The standard approach to the non-linear scales in $P(k)$ adopted in weak lensing analyses by DES  and the Kilo Degree Survey~\citep[KiDS,][]{Joudaki:2019pmv} is the \emph{Halo Model}~\citep[reviewed in e.g.][]{cooray2002halo}. The primary DES-Y1 analysis uses the halo model inspired fitting function \textsc{halofit}~\citep{2003MNRAS.341.1311S,2012MNRAS.420.2551B,Takahashi:2012em}. \textsc{halofit} is, however, not calibrated for use with ULA/FDM cosmologies~\citep{Hlozek:2016lzm}, and so a more physical approach is required. The halo model implementation \HMC~\citep{Mead_2015} provides such a model, which can be adapted to both warm DM (WDM) and FDM~\citep{Marsh:2016vgj}.~\footnote{\HMC~modifications in \cite{Marsh:2016vgj} accurately reproduce the WDM non-linear $P(k)$ ratio to CDM in \cite{Corasaniti:2016epp} to $\mathcal{O}(\rm few)\%$ accuracy. } \HMC\,is a complementary and alternative method to power spectrum emulators. It is useful for weak lensing, which is dominated by high density regions, where an analytical model saves considerable computational cost. Using the halo model, weak lensing bias parameters and other uncertainties in the model are easily accounted for in Bayesian parameter estimation, i.e. marginalised over.

%~\footnote{While this work was in preparation, \HMC~was updated \cite{Mead:2020vgs}, however we kept our analysis pipeline with the 2015 version.}

The halo model in \HMC\,is modified with physically motivated fitting parameters in order to match emulators and simulations of the non-linear power spectrum to within around 2\% accuracy. \HMC\,also includes a model for the important effect of baryonic feedback from Active Galactic Nuclei (AGNs), which affects $P(k)$ at the low redshifts $0\lesssim z \lesssim 2$ of the DES-Y1 galaxy sample. In the present analysis, following \cite{Abbott:2017wau}, we mask small scales in the DES-Y1 data that are affected by AGN feedback. We comment in our conclusions on the effect of including these scales in future analyses.

As described below, the halo model uses physical inputs of the linear matter power spectrum, the halo mass function (HMF), $n(M)$, and halo density profile parameterised by the halo concentration-mass relation, $c(M)$. This makes the halo model adaptable to different DM theories, and can account for physics observed in single simulations, but not performed in significant numbers to build an emulator around. \HMC\,further introduces a phenomenological model for the transition between the one halo and two halo terms in the quasi-linear regime, calibrated to $N$-body simulations with different initial transfer functions that cover a range of scales of interest for FDM.

In the following we introduce physically motivated modifications to \HMC\,based on observations from $N$-body simulations to model the non-linear scales of FDM. We then propagate (approximated) systematic uncertainties in the models adopted for $n(M)$, $c(M)$, and the quasi-linear smoothing, along with all relevant DES nuisance parameters. This allows us to perform a search for evidence of the FDM Jeans scale in the galaxy shear correlation function.

We furthermore perform a \emph{global fit} incorporating both DES and \emph{Planck} data, which are highly complementary in the cosmological parameter space. \emph{Planck} effectively anchors all the primary cosmological parameters, leaving the DES correlation functions as a lever arm into small scales, increasing the sensitivity to the FDM Jeans scale (and thus $m$) without any degeneracy with other parameters of the model.

We can anticipate our results by considering how weak lensing data can be used as a probe of the linear power spectrum, $P(k)$~\citep{Tegmark:2002cy}. As shown in \cite{Chabanier:2019eai} the \emph{Planck} CMB constrains $P(k)$ for wavevectors $k<0.3\, h\,\text{Mpc}^{-1}$, while the final DES bin covers $1\, h\,\text{Mpc}^{-1}<k<5\, h\,\text{Mpc}^{-1}$. The ULA power spectrum is suppressed below the Jeans scale at matter-radiation equality, given by~\citep{Hu:2000ke}: 
\begin{equation}
k_{\rm J,eq}=12.9 \, h\text{ Mpc}^{-1}\left(\frac{0.7}{h}\right)\left(\frac{m}{10^{-22}\text{ eV}}\right)^{1/2}\label{eqn:kj_eq}
\end{equation}
Thus, extending constraints to $P(k)$ by an order of magnitude in $k$ using DES compared to \emph{Planck} alone should leverage two orders of magnitude in sensitivity to the FDM mass. This improvement in the limit to $m$ driven by DES is evident in the one dimensional posterior derived from our analysis, shown in Fig.~\ref{fig:posterior_1d} (these results are discussed further in Section~\ref{sec:discussion}).

This paper is organised as follows. We start by introducing cosmic shear as a probe of dark matter and outline the halo model formalism including the relevant changes to the model in the case of FDM in Section~\ref{sec:observables}. We outline our statistical methodology and the DES-Y1 and \textit{Planck} data in Section~\ref{sec:data}. We present our results in Section~\ref{sec:dm_constraints} and conclude in Section~\ref{sec:discussion}. The Supplementary Material considers the introduction of the galaxy correlation function, which we omit from our main analysis due to uncertainty in the use of a scale independent galaxy bias, discusses aspects of the ULA halo model not included in our analysis, and discusses massive neutrinos in more detail.

\section{Observables and Models} \label{sec:observables}

\subsection{DES-Y1 Observables}
\label{sec:DES_observables}
We now introduce the observables analyzed in this study. We show how these observables depend on the total matter density power spectrum $P(k)$, which ultimately allows us to set constraints on the viable range of FDM particle mass $m$.

The DES-Y1 survey~\cite{Abbott:2017wau} measures the distribution of the number and ellipticites of galaxies, covering a total redshift range of $0\lesssim z \lesssim 2$. The data are divided in tomographic redshift bins $i$, where the number of bins is four or five, depending on the respective observable. Each redshift bin $i$ contains a distribution of galaxies $n^i(z)$, and the total number of galaxies in each bin is given $\bar n^i=\int {\rm d} z\, n^i(z)$.
The measured distributions of galaxies and their ellipticities yield a 
set of three two-point observables, which we refer to as `3x2pt' below. It consists of the individual observables galaxy clustering $w(\theta)$,
galaxy-galaxy lensing $\gamma_t(\theta)$, and the two components of the cosmic shear $\xi^\pm(\theta)$. In detail, $w(\theta)$ measures the distribution of pairs of angular separations of galaxies as compared to what is expected assuming a random distribution. The observables $\gamma_t(\theta)$ and $\xi_\pm(\theta)$ quantify shape distortions due to lensing by foreground mass distributions.  $\gamma_t(\theta)$ measures, for pairs of source and lens galaxies, the ellipticities of the the source galaxies tangential to the line connecting to the lens galaxy. $\xi_+(\theta)$ and $\xi_-(\theta)$, on the other hand, express for pairs of source galaxies the sum (+) and the difference (-) of the product of the tangential ellipticities  and the cross ellipticities. 

The effects measured by DES-Y1 are all sourced by the distribution of matter along the light of sight, and hence the corresponding 2pt observables are predicted via projections of the power spectrum $P(\ell,\chi)$ along this direction~\citep{Abbott:2017wau,krause2017dark}:
\begin{align}
    P^{ij}_{\kappa\kappa,\kappa g,gg}(\ell)=\int \rm{d}\chi\, \frac{q^i_{g/\kappa}q^j_{g/\kappa}}{\chi^2}\times P\left(\frac{\ell+1/2}{\chi},z(\chi)\right),\label{eq:projection}
\end{align}
where the weighting functions $q^i_{g/\kappa}$ are defined as
\begin{align}
    q^i_{g}(k,\chi)=b^i(k,z(\chi))\frac{n^i_g(z(\chi))}{\bar n^i_g}\frac{\rm dz}{\rm d \chi}\label{eq:galaxy_weight}
\end{align} and
\begin{align}
    q^i_\kappa(\chi)=\frac{3H_0^2\Omega_m}{2c^2}\frac{\chi}{a(\chi)}\int_\chi^{\chi_h}\,\frac{n^i_\kappa(z(\chi'))\rm d z/\rm d\chi'}{\bar n^i_\kappa}\frac{\chi'-\chi}{\chi'}
\rm d \chi'.\label{eq:lens_efficiency}
\end{align}
The comoving distance is $\chi$, and $b$ is the linear bias.
The final observables are calculated by transforming to real-space as follows:
\begin{align}
    w^i(\theta)&=\int\frac{\rm d \ell\,\ell}{2\pi} J_0(\theta \ell)P^{ij}_{gg}(\ell)\label{eq:w}\\
    \gamma^{ij}_t(\theta)&=(1+m^j)\int\frac{\rm d \ell\,\ell}{2\pi} J_2(\theta \ell)P^{ij}_{g\kappa}(\ell)\label{eq:gamma}\\
    \xi^{ij}_{+/-}(\theta)&=(1+m^i)(1+m^j)\int\frac{\rm d \ell\,\ell}{2\pi} J_{0/4}(\theta \ell)P^{ij}_{\kappa\kappa}(\ell),\label{eq:xi_pm}
\end{align}
where $J_\nu$ denotes the Bessel function of the first kind, and the parameter $m^i$ is the multiplicative shear bias.

When galaxies are used as tracers for the matter, we need the linear bias $b^i(k,z(\chi))$ to relate the distribution of galaxies to the underlying total matter distribution, resulting in the radial weight function Eq.~\eqref{eq:galaxy_weight}. By contrast, the lensing efficency Eq.~\eqref{eq:lens_efficiency} only depends on the matter density and the galaxy distribution along the line of sight. Consequently, the projection $P^{ij}_{\kappa g}(\ell)$ and hence the observable $\gamma_t$ depends linearly on the galaxy bias, while the projection $P^{ij}_{gg}(\ell)$ and hence the observable $w$ depends quadratically on the galaxy bias. Thus, measuring both $w$ and $\gamma_t$ allows the linear bias parameters to be measured~\citep{Abbott:2017wau}. 

%%%%%%%%%%%%%%%%%%%%%%%%%%%%%%%%%%%%%
% Shear figure
%%%%%%%%%%%%%%%%%%%%%%%%%%%%%%%%%%
\begin{figure*}
\begin{center}
 \includegraphics[width=.9
\textwidth]{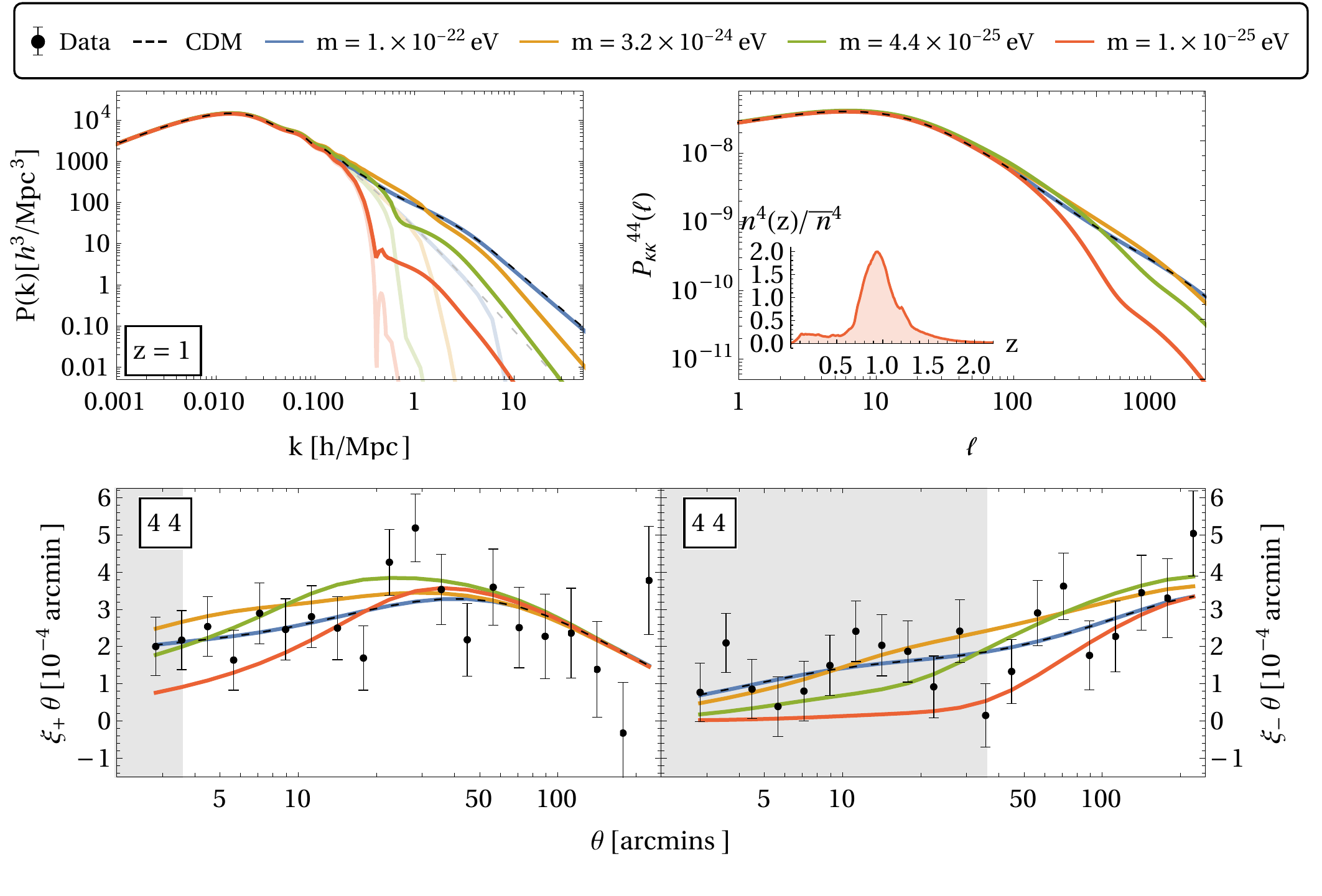}\\
 \caption{Cosmic shear data and model predictions for CDM and FDM with four different values for the mass parameter $m$. We show our model predictions for the linear (muted colours) and non-linear (saturated colours) power spectrum at redshift $z=1$ in the top left panel, the respective predictions for CDM are shown in dashed, black. The redshift is chosen near the peak of the galaxy distribution (inset top right pannel) in the fourth DES-Y1 redshift bin. We show the line-of-sight projected power spectrum for the $4,4$ bin in the top right panel, with the galaxy distribution used in the integral kernel for the projection inset. Finally we compare our predictions for the shear observables $\xi_+$ (left) and $\xi_-$ (right) to the data in the $4-4$ bin in the bottom row. We mark DES-Y1 data points in black, indicating the square-roots of the diagonal entries of the covariance matrix as vertical bars. The gray shaded regions mask angular scales excluded from the analysis due to modeling uncertainties. The case $m=10^{-22}\text{ eV}$ is indistinguishable from CDM over the scales shown. Our model predictions compared to the full DES-Y1 data are shown in the Supplementary Material.
}
 \label{fig:shear_powerspectra}
\end{center}
\end{figure*}

%%%%%%%%%%%%%%%%%%%%%%%%%%%%%%%%%%%%%%

In FDM models, the bias parameter is expected to have additional scale dependence due to the Jeans scale~\citep{Hlozek:2014lca}. Recently \cite{Lague:2021frh} simulated Baryon Oscillation Spectroscopic Survey (BOSS) data in cosmologies with a sub-dominant fraction of ULAs with $m\leq 10^{-24}\text{ eV}$, modelling the galaxy bias in order to derive rigorous constraints. Such an analysis for DES, with heavy ULAs composing all the DM, would require fully non-linear simulations and is beyond the scope of the present work. However, we expect that for the relatively heavy ULAs considered in the present work ($m\gtrsim 10^{-24}\text{ eV}$ consistent with CMB and BOSS lower bounds) the galaxy bias for ULAs will be linear and (almost) scale-independent, just like the CDM galaxy bias. Nonetheless, given our relative ignorance of how to model the galaxy bias for ULAs in our mass range of interest, we consider the shear only analysis (using only $\xi_\pm$), to be the most conservative, and present it as our main analysis. 

We show the cosmic shear data together with the predictions for CDM as well as for four different FDM masses $m$ for our fiducial cosmology for the $4,4$ bin in Fig.~\ref{fig:shear_powerspectra}. The full set of bin combinations is illustrated in 
Fig.~\ref{fig:shear_plusmminus_models_data}.
The full 3x2pt analysis is considered in the Supplementary Material, where we show how the respective observables $\gamma_t$ and $w$ are affected by FDM (Fig.~\ref{fig:extra_models_data}). As anticipated in the introduction, for masses $\lesssim 0.5\times10^{-23}\rm{eV}$, the predictions from FDM model differ visibly from the CDM case.

For the cosmic shear observables $\xi_\pm$, the angular scales $\theta$ measured by DES-Y1 and shown in Fig.~\ref{fig:shear_plusmminus_models_data} are related to the radial distances $k$ in the power spectrum $P(k)$ through the projection Eq.~\ref{eq:projection} and the filtering with the Bessel functions $J_{0/4}$ in Eq.~\eqref{eq:xi_pm}. As noted above, we can use the results by \cite{Chabanier:2019eai} to deduce that the DES-Y1 data are sensitive to values of $k\lesssim5\,h\mathrm{Mpc}^{-1}$ and hence to the quasi-linear and mildly non-linear scales of the power spectrum $P(k)$. As we discuss below, this sensitivity makes the DES-Y1 data set an interesting probe of DM. On the other hand, it requires modelling of the power spectrum beyond linear theory.

\subsection{The FDM Power Spectrum}\label{sec:DES_and_DM}

Because the DES-Y1 observables defined by Eqs.~\eqref{eq:w}, \eqref{eq:gamma} and \eqref{eq:xi_pm} are related to the total matter density power spectrum, they are all sensitive to the distribution of DM. As motivated above, in our analysis, we focus on cosmic shear, which depends on the total power spectrum of the gravitational potential. The gravitational potential is related to the total matter density by the Poisson equation. It is an unbiased tracer of the total matter content of the Universe, including DM. Since the power spectrum measures the clustering of DM under gravity it is sensitive to any departures from CDM. 

In the case of FDM, these departures come at different scales. At linear scales, we predict two different effects. Firstly, the background energy density evolves differently. For the scales of interest in the present study, $m\gg 10^{-27}\text{ eV}$, this effect does not play a role~\citep{Hlozek:2014lca}. Secondly, gradients in the FDM field play the role of pressure, and lead to a Jeans scale suppressing structure formation~\citep{Khlopov:1985jw}. Both of these effects are captured by solving the perturbed Klein-Gordon-Einstein equations, coupled to the rest of the contents of the Universe via the metric and the Einstein equations. 
In the present work, the linear power spectrum is evaluated with \textsc{camb}~\citep{Lewis:1999bs}, modified in the case of ULAs/FDM to \acamb~\citep{Hlozek:2014lca,acamb_link}.~\footnote{\acamb is based on the 2013 release of \textsc{camb}. We checked via a visual comparison of the posterior distributions that our results in the case of CDM are unchanged between the 2013 version of \textsc{camb} and the more recent versions available within \cosmosis. Work on an updated version of \acamb is ongoing.} \acamb solves the Klein-Gordon equation in the Madelung fluid variables, using a WKB approximation for the sound speed at late times.

The suppression of the FDM linear power with respect to the CDM case congruously leads to a different distribution of halos, in particular a reduction in the number and inner density of halos near to and below the Jeans mass. 
We implement these effects on the non-linear power spectrum within a modified version of the Halo Model as described in Section~\ref{subsec:axionHM}.
Finally, the specific properties of FDM leads to wave-like effects on very small scales, in particular the formation of solitonic cores in the centers of halos. Since these effects are restricted to scales far below the scales probed by the DES-Y1 data sets, we do not discuss them in the context of this analysis. The justification to neglect solitons and certain other effects in our analysis are discussed in detail in the Supplementary Material.
%%%%%%%%%%%%%%%%%%%%%%%%%%%%%%%%%%%%%%%%%%%%%%%%%%%%%%%%%%%%%%%%

\subsection{Halo Model and \HMC}
\label{sec:nl_power}
The starting point of all cosmological analyses is linear perturbation theory. Beyond the linear regime, different variants of the Halo Model provide a semi-analytical construction of the non-linear power spectrum from the corresponding power spectrum obtained from linear theory.
We summarise the general idea behind this approach below. We furthermore give a short description of \HMC~\citep{Mead_2015}, the specific implementation of the Halo Model which we use as basis for our adaption of the Halo Model. In a final step, we specify the modelling choices we made for our ULA/FDM version of \HMC\,in S ec.~\ref{subsec:axionHM}.

%%%%%%%%%%%%%%%%%%%%%%%%%%%%%%%%%%%%%%%%%%%%%%%%%%%%%%%%%%%%%%%%%%%%%%%%%%%%%%%%%
\subsubsection{Halo Model -- Basic Concept}
\label{subsec:HM}
%%%%%%%%%%%%%%%%%%%%%%%%%%%%%%%%%%%%%%%%%%%%%%%%%%%%%%%%%%%%%%%%%%%%%%%%%%%%%%%%%
Calculations in cosmological models are often more conveniently done in $k$ space and hence we usually define observables in terms of the power spectrum $P(k)$. However, the Halo Model is based on physical concepts in real space like the halo density profile and the corresponding halo mass contained within a certain radius. We therefore find it more favourable to introduce the Halo Model in terms of the matter density correlation function $\xi(\vec x_1,\vec x_2)$, the Fourier transform of the power spectrum $P(k)$

Assuming that all matter is contained with halos of density $\rho(\vec x)$, we can factorise $\xi(\vec x_1,\vec x_2)$ as~\citep{cooray2002halo,hayashi2008understanding}
\begin{align}
 \xi(\vec x_1,\vec x_2)&=\langle\delta(\vec x_1),\delta(\vec x_2)\rangle \nonumber \\
  &=\langle \frac{\rho_{h}(\vec x_1)}{\bar\rho}\frac{\rho_{h'}(\vec x_2)}{\bar\rho}\rangle+\langle \left(\frac{\rho_{h}(\vec x_1)-\bar\rho}{\bar\rho}\right)\left(\frac{\rho_{h}(\vec x_2)-\bar\rho}{\bar\rho}\right)\rangle\nonumber\\
 &\equiv\xi^{2h}(\vec x_1,\vec x_2)+\xi^{1h}(\vec x_1,\vec x_2),\label{eq:def_1h_2h}
\end{align}
where we define the two-halo term $\xi^{2h}$ as the correlation function between densities attributed to two different halos $h$, $h'$, while the one-halo term $\xi^{1h}$ is defined as the correlation between densities attributed to the same halo. It is commonly assumed that the density distribution within a halo, $\rho_h$, is a function of space with the total mass $M$ of the halo as a single parameter, i.e. $\rho_h=\rho_h(\vec x|M)$. The ensemble average in Eq.~\eqref{eq:def_1h_2h} can be replaced by the average over space and the average over the halo mass $M$, as follows~\citep{cooray2002halo}
\begin{align}
 \xi^{1h}(\vec x_1,\vec x_2)&=\frac{1}{\bar\rho^2}\int \mathrm{d}M\,n(M)\times\nonumber\\
 \int \mathrm d \vec r\, &\left(\rho_h(\vec x_1-\vec r|M)-\bar\rho\right)\left(\rho_h(\vec x_2-\vec r|M)-\bar\rho\right)\label{eq:1h}\\
 \xi^{2h}(\vec x_1,\vec x_2)&=\frac{1}{\bar\rho^2}\int \mathrm{d}M_1\,n(M_1)\int \mathrm{d}M_2\,n(M_2)\,\times\nonumber\\
 \int\!\int\mathrm d \vec r_1\mathrm d \vec r_2\, &\rho_{h_1}(\vec x_1-\vec r_1|M_1)\rho_{h_2}(\vec x_2-\vec r_2|M_2)\,\xi_{hh}(\vec r_1,\vec r_2|M_1,M_2).\label{eq:2h}
\end{align}
In the above equation, $n(M)$ denotes the halo mass function, which describes the number of halos per unit volume, as a function of their total mass $M$. The term $\xi_{hh}(\vec r_1,\vec r_2 | \allowbreak M_1,M_2)$ encodes the correlation between two halos of mass $M_1$ at point $\vec r_1$ and mass $M_2$ at point $\vec r_2$, respectively. If the correlation between halos varies slowly on scales of the order of the halo sizes, the halo density distributions $\rho_{h_1}$, $\rho_{h_2}$ in Eq.~\eqref{eq:2h} can be approximated by Dirac-$\delta$-functions times the respective masses, i.e. $\rho_{h}(\vec x-\vec r|M)\sim M\,\delta^D(\vec x-\vec r)$. The correlation function $\xi_{hh}$ itself can as first approximation be replaced by the correlation function $\xi_{lin}$ obtained from linear theory, times the linear bias $b(M)$
\begin{align}
  \xi^{2h}(\vec x_1,\vec x_2)\sim b(M_1)b(M_2)\xi^{lin}(\vec x_1,\vec x_2)\sim \xi^{lin}(\vec x_1,\vec x_2)\label{eq:Approx2H}.
\end{align}
which is justified at large scales $|\vec x_1-\vec x_2|\gg 1$ because at large scales we can again assume $\rho_{h}(\vec x-\vec r|M)\sim M\,\delta^D(\vec x-\vec r)$, and $\int \mathrm{d}M\,n(M)\,b(M)\,M=\bar\rho$ by definition of the linear bias. At smaller scales, in principal we need to account for the linear bias $b(M)$. However, at these scales, the total correlation function is dominated by the one-halo term (this approximation as implemented in \HMC\,is discussed further below).

Hence, within the framework of the halo model introduced above, two components need to be predicted: the halo mass function $n(M)$ and the halo density profile $\rho_h(\vec r|M)$. A conventional choice for the halo mass function is obtained from the ellipsoidal collapse model~\citep{bond1996peak,Sheth:1999su,Sheth:1999mn,cooray2002halo} within the ~\cite{press187formation} approach. The result is
\begin{align}
 n^{\rm ST}(M)=&\frac{\bar\rho}{M}f(\nu)\frac{\mathrm d \nu}{\mathrm dM}\, , \label{eqn:hmf_definition}\\
 f(\nu)=&A\left[1+\frac{1}{(a\,\nu^2)^p}\right]e^{-a\,\nu^2/2} \label{eq:Sheth-Tormen}
\end{align}
where the parameters $a=0.707$, $p=0.3$ and $A=0.2162$ are obtained from empirical fits to $\Lambda$CDM simulations~\citep{Mead_2015}. The variable $\nu$ is defined by the ratio
\begin{align}
 \nu&\equiv\frac{\delta_c}{\sigma}.
\end{align}
In the above equation, the variable $\delta_c$ denotes the critical density defined as the minimum density for which a region collapses under its own gravity. The variable $\sigma^2$, defined by
\begin{align}
 \sigma^2(M)=\langle\delta_s(M)^2\rangle, \label{eqn:def_sigma}
\end{align}
denotes the autocorrelation function measuring the correlation between smoothed overdensities $\delta_s(\vec r,M)$, which are averaged over a sphere with radius $R$ as follows:
\begin{align}
 \delta_s(\vec r,M)=\frac{3}{4\pi R(M)^3}\int_{0}^{\vec R}\mathrm d^3 x\,\delta(\vec x-\vec r),
\end{align}
where the smoothing radius $R=R(M)$ is chosen such that the corresponding sphere contains on average the mass $M=4\pi/3\bar\rho R^3$~\citep{mo2010galaxy}.

The conventional model for the second component of the halo model, the halo density profile, $\rho_h(\vec r|M)$, is the  Navarro-Frenk-White (NFW) profile, defined as~\citep{navarro1997universal,Understanding_halo_mass}
\begin{align}
 \rho_h(r|\rho_s,r_s)-\bar\rho=\frac{\rho_s}{(r/r_s)(1+r/r_s)^2},\label{eq:NFW}
\end{align}
which depends on the two parameters scale density $\rho_s$ and scale radius $r_s$. In order to obtain a finite halo mass $M=\int \rho_h(r) d^3r$, the NFW profile, Eq.~\eqref{eq:NFW} needs to be truncated. Conventionally, the truncation radius is set to the virial radius $R_{vir}$. Defining the concentration parameter $c$ as
\begin{align}
   c\equiv\frac{R_{vir}}{r_s},
\end{align}
and noting 
\begin{align}
  \rho_s=\frac{c^3 M}{4 \pi  R_{vir}^3\left(
\log(1+c)-c/(1+c)\right)},  
\end{align}
we can express the NFW profile in terms of the variables $M$, $R_{vir}$  and $c$ as follows~\citep{cooray2002halo}
\begin{align}
 \rho_h(r|\rho_s,r_s)\bigg|_{r<R_{vir}}\longrightarrow\;\rho_h(r|M,R_{vir},c)\bigg|_{r<R_{vir}}.
\end{align}
With the halo density profile and the halo mass function defined, we have constructed a general halo model.

\subsubsection{Implementation in \HMC}
We now describe \HMC, an implementation of the Halo Model by \cite{Mead_2015}. The authors match the Halo Model parameters to a set of simulations with varying cosmological parameters \cite{cosmic_emu}. Furthermore, to improve the predictive power of the model, they introduce a range of additional parameters which are fitted to these simulations. A summary of all  parameters as well as the best fit values reported by~\cite{Mead_2015} is given in Table~\ref{tab:mead2015}. Below, we explain the physical motivation as well as the the effect of each parameter on the power spectrum.~\footnote{While this work was in preparation, \HMC~was updated \cite{Mead:2020vgs}, however we kept our analysis pipeline with the 2015 version.}

For the Halo Model parameters halo mass $M$,  concentration parameter $c$, and critical density $\delta_c$, the authors use the following prescriptions: 
The halo mass parameter $M$ is defined such that the average density of the halo $\bar\rho_h(M)$ is greater than the background density $\bar\rho$ by a factor $\Delta_v=\bar\rho_h/\bar\rho$.  $\Delta_v$ is fitted to simulations according to the prescription in Table~\ref{tab:mead2015}.

The concentration parameter $c$ is calculated according to the prescription of \cite{bullock_2001}
\begin{equation}
c^{\rm B}(M,z,P^{lin}(k)) = A \frac{1+z_{\rm coll}(M,P^{lin}(k))}{1+z},    \label{eq:c_of_m}
\end{equation}
where the minimum halo concentration parameter $A$ in the above equation is again fitted to simulations~\citep{Mead_2015}. The redshift $z_{\rm coll}(M)$ of collapse for halos of mass $M$ is defined as the redshift where the mass of the halo was lighter by a factor of $f_{\rm coll}=0.01$ as compared to the current mass at redshift $z$, estimated according to the relation
\begin{align}
    \frac{g(z_{\rm coll})}{g(z)}\sigma(f_{\rm coll}M,P^{lin}(k))=\delta_c,\label{eq:z_coll}
\end{align}
where $g$ is the linear growth factor. Thus, the concentration-mass relation $c(M,z)$ is determined by the input linear theory power spectrum that is used to compute $g$ and $\sigma$, which we denote explicitly in Eqs.~\eqref{eq:c_of_m}) and \eqref{eq:z_coll}). The critical density for collapse, $\delta_c$, has the functional form given in Table~\ref{tab:mead2015}, with co-efficients determined by best fit to simulations.

In addition to tuning the generic Halo Model parameters $M$, $c$ and $\delta_c$ to simulations, \cite{Mead_2015} introduce a range of modifications, which are intended to parametrize physical effects which are neglected in the basic Halo Model introduced in Section~\ref{subsec:HM}. These modifications affect the one-halo term, the two-halo term and the combination of both terms as described below. 

\HMC\,approximates the two-halo term as given in Eq.~\eqref{eq:Approx2H}, neglecting the bias. This argument is confirmed by \cite{Mead:2020vgs}, who find that neglecting the linear bias, $b(M)$, in Eq.~\eqref{eq:Approx2H} only changes the total correlation function at a significant level (compared to the desired accuracy) once the one halo term is already dominant. Thus, $b(M)$, can be neglected within the desired accuracy of \HMC. In fact, this approximation over-predicts the correlation function on smaller scales of the order of the halo radii. Hence, \cite{Mead_2015} adopt an empirical damping model as follows
\begin{align}
 \Delta^2_{2h}(k)&=\left[1-f\tanh^2(k\sigma_v/\sqrt{f})\right]\Delta_{lin}^2(k),\label{eq:mead2h}\\
 \sigma^2_v&=\frac{1}{3}\int_0^\infty\frac{\Delta_{lin}^2(k)}{k^3}\mathrm dk,
\end{align}
where $f$ fixed by an empirical fit, and $\Delta^2_{lin}(k)=4\pi V(k/2\pi)^3P(k)$, where the power spectrum $P(k)$ is the Fourier transform of the correlation function $\xi(x)$.

For the one-halo term, \cite{Mead_2015} introduce two modifications.
The first modification is supposed to correct for the problem that the simple expression Eq.~\eqref{eq:1h} does respect that halos are mutually exclusive, which results in too large contributions of the one-halo term to the correlation function at very large scales. The correction is given by
\begin{align}
 \Delta^2_{1h}=\left[1-e^{-(k/k_*)^2}\right]\Delta^2_{1h,gen},\label{eq:mead1h}
\end{align}
where $\Delta^2_{1h,gen}=4\pi V(k/2\pi)^3P_{1h,gen}(k)$, with $P_{1h,gen}(k)$ being the Fourier transform of the generic one-halo term $\xi^{1h}(r)$  defined in Eq.~\eqref{eq:1h}. The parameter $k_*$ is again fixed by an empirical fit. The authors furthermore modify the halo density profile in Fourier space:
\begin{align}
 \rho_h(k|M,R_{vir})\longrightarrow\;\rho(\nu^\eta k,M,R_{vir})\label{eq:hm_halo_trafo}
\end{align}
where $\eta$ is again a heuristic fit parameter. With this modification of the density profile, the total halo mass $M$ is unaltered. Furthermore, the profile of halos with $\nu=1$ does not change, while for halos with $\nu>1$ ($\nu<1$) the profile flattens (sharpens), as $\rho_s$ decreases (increases) with $\nu^{-3\eta}$ while the scale radius $r_s$ and the cut-off increase (decrease) as $r_s\,\nu^\eta$ and $R_{vir}\,\nu^\eta$, respectively. 

Another shortcoming of the generic halo model is the crude transition between the one-halo term appropriate for small scales and the two-halo term appropriate for large scales. Physically, one expects in the so-called quasi-linear regime a smooth transition. \cite{Mead_2015} fit this smooth transition regime with the following functional form
\begin{align}
 \Delta^2(k)=\left[(\Delta^2_{2h})^\alpha+(\Delta^2_{1h})\right]^{1/\alpha},
 \label{eqn:one-to-two_smooth}
\end{align}
with $\alpha$ an empirical fit parameter given in Table~\ref{tab:mead2015}, and $\Delta^2_{2h}$ and $\Delta^2_{1h}$ defined according to Eqs.~\eqref{eq:mead2h} and \eqref{eq:mead1h}, respectively. The fit parameter $\alpha$ is determined by the effective spectral index, $n_{\rm eff}$, given by:
\begin{equation}
    3+n_{\rm eff}(z)=-\left.\frac{d\ln \sigma^2(R,z)}{d\ln R}\right|_{R=R_{\rm nl}}\, ,
    \label{eqn:neff_definition}
\end{equation}
where $\sigma^2$ is the variance of the linear power spectrum, and the non-linear scale $R_{\rm nl}$ is defined by $\sigma^2(R_{\rm nl})=1$. We discuss the smoothing Eq.~\eqref{eqn:one-to-two_smooth} and its role in FDM models in depth in Section~\ref{sec:smoothing}.

Finally, \cite{Mead_2015} fit the critical density $\delta_c$ the virialised overdensity $\Delta_v=3m/4\pi R_{vir}^3\bar\rho$ used to define $R_{vir}$, as well as the parameter $A$ used to define the concentration parameter $c(A)$. See Table~\ref{tab:mead2015}.

\begin{table*}
    \centering
    \begin{tabular}{c|c|c}
    Parameter & Description & \cite{Mead_2015} fit value\\
    \hline
    $\Delta_v$  & Virialised halo overdensity& 418$\,\Omega_m(z)^{-0.352}$  \\
    $\delta_c$& Linear collapse threshold&$1.59+0.0314\ln{\sigma_8(z)}$\\
    $\eta$&Halo bloating &$0.603-0.3\sigma_8(z)$\\
    $f$&Linear spectrum transition damping&$0.199\sigma_8(z)^{4.29}$\\
$k_*$& One-halo damping wavenumber&$0.584\sigma_v(z)^{-1}$\\
$A$& Min. halo concentration&$3.13$\\
$\alpha$& Quasi-linear one- to two-halo term softening&$2.93\times1.77^{n_\mathrm{eff}}$\\
\hline
    \end{tabular}
    \caption{The halo model parameters and fitted values from \protect\cite{Mead_2015}. The additional non-linear nuisance parameters varied in this work are specified in Eq.~\eqref{eqn:non-linear_nuisance}.}
    %\caption{The halo model parameters and fitted values from. The additional non-linear nuisance parameters varied in this work are specified in Eq.~\eqref{eqn:non-linear_nuisance}.}    
    \label{tab:mead2015}
\end{table*}

%%%%%%%%%%%%%%%%%%%%%%%%%%%%%%%%%%%%%%%%%%%%%%%%%%%%%%%%%%%%%%%%%%%%%%%%%%%%%%%%%
%%%%%%%%%%%%%%%%%%%%%%%%%%%%%%%%%%%%%%%%%%%%%%%%%%%%%%%%%%%%%%%%%%%%%%%%%%%%%%%%%

%%%%%%%%%%%%%%%%%%%%%%%%%%%%%%%%%%%%%%%%%%%%%%%%%%%%%%%%%%%%%%%%%%%%%%%%%%%%%%%%%
\subsection{Halo Model and Lensing Observables for FDM}\label{subsec:axionHM}

The lensing observables for FDM are shown in Figs.~\ref{fig:shear_plusmminus_models_data} and~\ref{fig:extra_models_data}. In the following we describe the modifications to the halo model that lead to the observed effects, which cause ULAs to be observationally distinct from CDM. Effects we did not include in our halo model, and which are deemed to be unimportant for DES-Y1 observables, are discussed in the Supplementary Material.

\subsubsection{Halo Mass Function and Concentration}

It is known that for models with a truncated linear power spectrum, $P^{lin}(k)$, such as models with FDM, but also with warm dark matter (WDM), the standard predictions for $n(M,P^{lin}(k))$ and $c(M,P^{lin}(k))$ described above disagree with the results of $N$-body simulations when using a real-space top-hat window function. Specifically, in simulations $n(M)$ and $c(M)$ display additional suppression and a cut-off compared to the predictions from the real space spherically averaged variance $\sigma(M,P^{lin}(k))$ computed from the linear power spectrum~\citep[e.g][]{2012MNRAS.424..684S,Schneider:2013ria,Schive:2015kza,Corasaniti:2016epp}. The additional suppression in $n(M)$ in an $N$-body simulation is visible after removing numerical artefacts \citep[``spurious haloes''][]{Wang:2007he}. However, such a suppression is expected from basic physical principles applied to models with a truncated linear power spectrum: we expect there to be no haloes below some cut-off scale, and a component of unbound DM. Similarly, a turnover in $c(M)$ at low $M$ can be interpreted as a turnover in the redshift of collapse, with low mass halos formed due to fragmentation of larger ones (see Eq.~\ref{eq:c_of_m}). \HMC\,imposes a minimum value of $c=A>1$ at large masses, and we keep the same minimum value also at low masses, implying $z_{\rm coll}>0$ for all halos. This is consistent with the assumption that all halos are described by NFW profiles, even if they are rare with low $n(M)$. 

The exact shape of $n(M)$ and $c(M)$ below the cut-off is not well known, since the spurious haloes are sometimes removed in a somewhat \emph{ad hoc} way. This motivates exploring a range of models for the cut-offs, and marginalising over this uncertainty when deriving cosmological constraints. We retain the real space  window function since it allows physical mass assignment, and is already implemented in \HMC. We model the cut-offs in $n(M)$ and $c(M)$ within \HMC\,building on the results reported from $N$-body simulations, and \cite{Marsh:2016vgj}. 

\begin{figure}
\begin{center}
 \includegraphics[width=.45\textwidth]{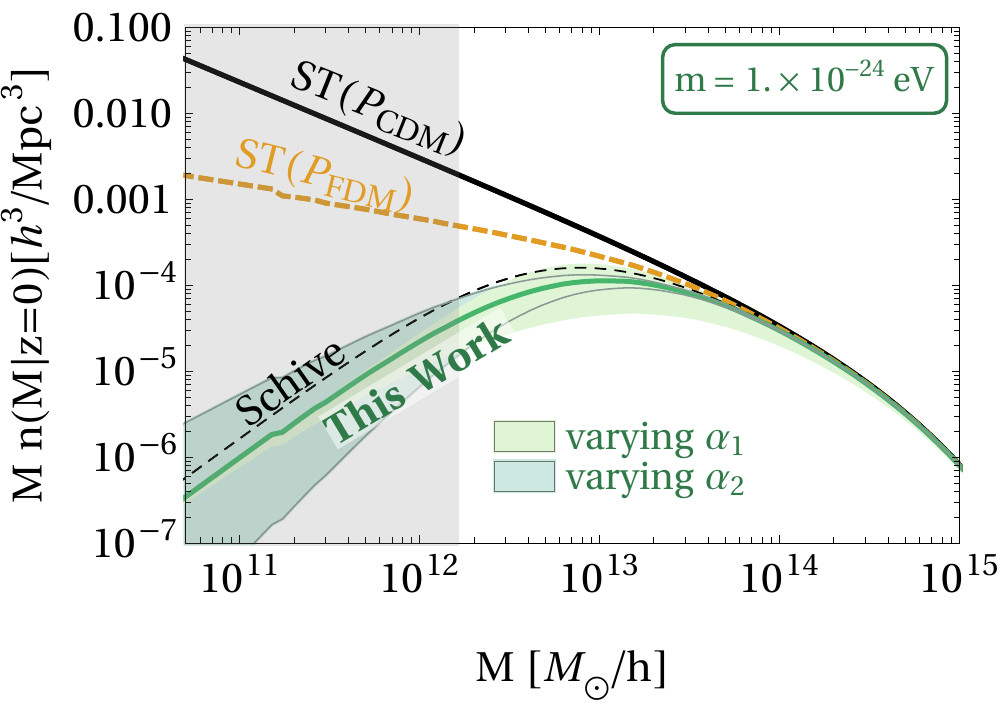}\\
\end{center}
\caption{Suppression of the Halo Mass Function in our model. We show the Sheth-Tormen prediction Eq.~\ref{eqn:hmf_definition} \protect\citep{Sheth:1999mn} based on a CDM (black, solid) and a FDM (orange, dashed) linear power spectrum, for a particle mass $m=10^{-24}$\,eV. In simulations, an additional suppression of the HMF is observed, parametrized by~\protect\cite{Schive:2015kza} (thin black, dashed). Our model of the additional suppression is shown in green. To account for modeling uncertainties we introduce two nuissance parameters $\alpha_1$, $\alpha_2$ (cf. Eq.~\ref{eq:corr_HMF_ULA}). We show the effect of varying $\alpha_1$ ($\alpha_2$), keeping $\alpha_2$ ($\alpha_1$) fixed in light green (dark green). The grey-shaded regions correspond to Halo Masses whose virial radius cannot be resolved on the scales present in the DES-Y1 data. Note that we pick different values for the particle mass parameter $m$ in Fig.~\ref{fig:nM_model} and Fig.~\ref{fig:cM_model}, because differences between FDM and CDM affect the HMF and the concentration mass relation at different scales.}
\label{fig:nM_model}
\end{figure}

In Fig.~\ref{fig:nM_model} we show different model predictions for the HMF. In solid, black, we show the Sheth-Tormen model $n^{\rm ST}(P_{\rm CDM})$, introduced in Eq.~\eqref{eq:Sheth-Tormen}, for a CDM linear power spectrum. In dashed, blue we show the Sheth-Tormen model $n^{\rm ST}(P_{\rm FDM})$ for a FDM linear power spectrum. Below some halo mass, the Sheth-Tormen model predicts a suppression of the $n^{\rm ST}(P_{\rm FDM})$ with respect to $n^{\rm ST}(P_{\rm CDM})$, which is rather uniform for varying $m$, reflecting the suppression of the linear power spectrum. However, as discussed above, the prediction from the Sheth-Tormen model, $n^{\rm ST}(P_{\rm FDM})$, deviates from simulation results which feature an additional suppression of the HMF. To assess the impact on prediction of the observables, we estimate the maximum visible halo masses $M$. Assuming spherical collapse with a critical mass $\delta_c$ as implemented in \HMC, we relate the the halo mass $M$ to a virial radius $R_{\rm vir}$, which can be converted to an angular scale via the angular diameter distance $D_A$.  To make a conservative estimate we take the redshift of $z=0.09$, below which are only 1\% of the DES-Y1 galaxy sample in the nearest redshift bin. We observe that differences between the CDM and FDM HMF need to occur for relatively large halo masses $M$ to impact the DES-Y1 observables, implying that this effect will be only relevant for very low particle masses. For consistency, we aim at matching the cut-off in the HMF and in the concentration mass relation to be discussed below to simulation results.
\cite{Schive:2015kza} showed that the suppression with respect to the CDM case can be described quite accurately by the following two-parameter model
\begin{align}
  n(P_{\rm CDM})=n^{\rm ST}(P_{\rm CDM})\,\Delta_n^{\rm CDM} (M_0,\alpha_1,\alpha_2),
\end{align}
where $n(P_{\rm CDM})$ is the HMF taking into account the additional suppression observed in simulations and $n^{\rm ST}(P_{\rm CDM})$ is the original HMF for CDM in the Sheth-Tormen model. The correction function with respect to the CDM case $n^{\rm ST}(P_{\rm CDM})$ is defined as
\begin{align}
    \Delta_n^{\rm CDM} (M^{(n)}_0,\alpha_1,\alpha_2)=\left[1+\left(\frac{M}{M^{(n)}_0}\right)^{-\alpha_1}\right]^{-(\alpha_2/\alpha_1)}
    \label{eq:corr_HMF_CDM}
\end{align}
with a scaling mass $M^{(n)}_0=1.6\times10^{10} (m_{Ax}/10^{-22}\rm{eV})^{-4/3} M_\odot$. For halo masses $M\gtrsim M^{(n)}_0$ the HMF agrees for FDM and CDM. For halo masses $M\lesssim M^{(n)}_0$, the FDM HMF is suppressed with respect to the CDM case. The steepness of the suppression is controlled by the parameter $\alpha_2$ and the sharpness of the transition at $M\sim M_0$ is controlled by $\alpha_1$. Fitting this model to $N$-body simulations results in $\alpha_1=1.1$, $\alpha_2/\alpha_1=2.2$~\citep{Schive:2015kza}. We show this model in thin, black dashed in Fig. \ref{fig:nM_model}.

To implement the additional suppression into \HMC, we want to correct the Sheth-Tormen model for FDM, $n^{\rm ST}(P_{FDM})$. We use the ansatz
\begin{align}
  n(M,P_{\rm FDM}) =n^{\rm ST}(P_{\rm FDM})\Delta_n^{\rm FDM} (M^{(n)}_0,\alpha_1,\alpha_2),
  \label{eq:corr_HMF_ULA}
\end{align}
where we choose the correction function to be of the same functional form as Eq.~\eqref{eq:corr_HMF_CDM}, and $n^{\rm ST}(P_{\rm FDM})$ is the Sheth-Tormen mass function computed with the FDM linear power spectrum. This is justified because for small halo masses $M\lesssim M^{(n)}_0$, the Sheth-Tormen prediction for both, the CDM and the ULA case are well described by a power-law. Re-fitting the slope-parameter $\alpha_2$ within our fiducial model, we find $\alpha_2=1.86$ for our correction function in Eq.~\eqref{eq:corr_HMF_ULA}. We show our model in solid, green in Fig.~\ref{fig:nM_model}.
%%%%%%%%%%%%%%%%%%%%%%%%%%%%%%%%%%%%%%%%%%%%%%%%%%%%%%%%%%%%%%%%%%%%%%%%%%%%%%%%%
%
%
% % 
\begin{figure}
 \begin{center}
 \includegraphics[width=.4\textwidth]{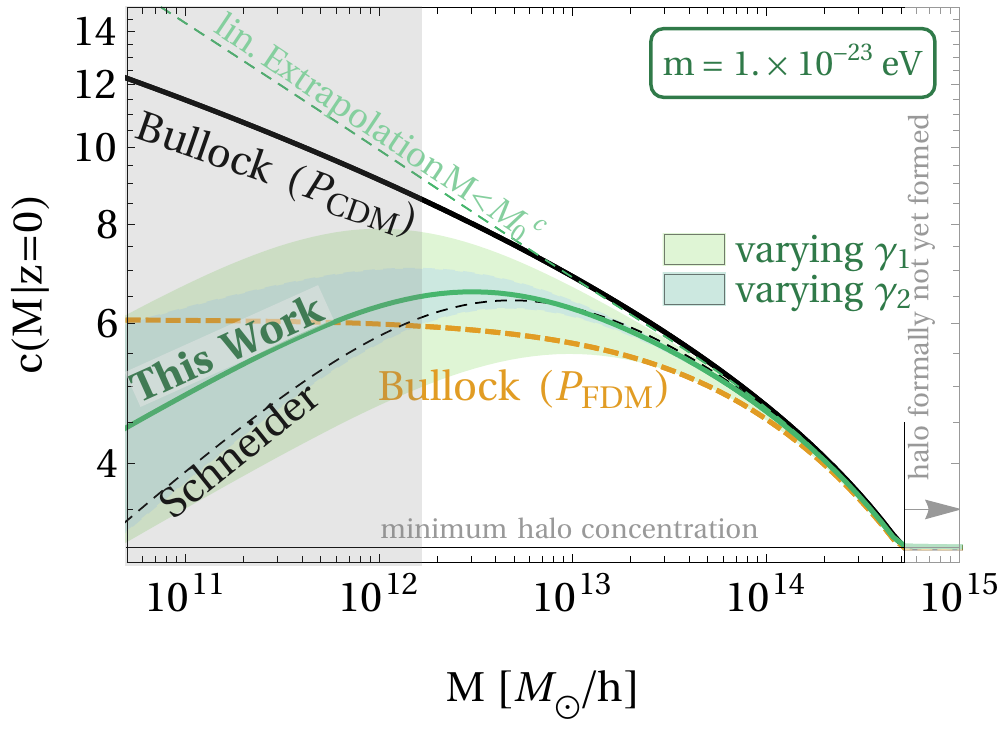}
\end{center}
\caption{Suppression of the concentration mass relation in our model. We show the case for CDM and FDM directly from the linear power spectrum using the \protect\cite{bullock_2001} relation
in \HMC, Eq.~\eqref{eq:c_of_m}. \protect\cite{Schneider:2013ria} found additional suppression below the half-mode mass in related WDM models. We include this extra suppression using two nuisance parameters, $\gamma_1$ and $\gamma_2$. We show the effects within each prior, holding the other one fixed, by the light green ($\gamma_1$) and dark green ($\gamma_2$) shaded regions. We denote the mass scale $M$ above which formally halos have not yet formed, i.e. $z_{\rm coll}<0$. For these halos, which are very rare according to the respective HMF, we assume the minimum halo concentration $A$.
The vertical shaded region indicates approximately halos that are below the DES resolution.
Note that we pick different values for the particle mass parameter $m$ in Fig.~\ref{fig:nM_model} and Fig.~\ref{fig:cM_model} to highlight the differences between FDM and CDM affect the HMF and the concentration mass relation, which occur on different scales.
}
\label{fig:cM_model}
\end{figure}

%%%%%%%%%%%%%%%%%%%%%%%%%%%%%%%%%%%%%%%%%%

Our approach for the concentration parameter is very similar to our treatment of the HMF. In Fig.~\ref{fig:cM_model}, we illustrate the different models for the concentration parameter. In solid, black, we show the \cite{bullock_2001} prediction, as introduced in Eq.~\eqref{eq:c_of_m}, for a CDM linear power spectrum, $c^{\rm B}(P_{\rm CDM})$. In thick blue, dashed we show the \cite{bullock_2001} prediction for the FDM case, $c^{\rm B}(P_{\rm CDM})$, with the modified collapse redshift, $z_{\rm coll}(P_{\rm FDM})$ computed from the FDM linear theory power spectrum from \acamb. As in the case of the HMF, $c^{\rm B}(P_{\rm FDM})$ is suppressed with respect to $c^{\rm B}(P_{\rm CDM})$, turning flat for small halo masses $M$. However, the suppression predicted by the \cite{bullock_2001} model is less than measured in simulations. Note that we show our models for the concentration parameter for a larger particle mass $m=10^{-23}$\,eV as compared to Fig.~\ref{fig:cM_model}. This is because the differences between FDM and CDM affect the concentration parameter at higher mass scales $M$, because the halo density profile, and hence the concentration parameter, trace the matter density at the time of collapse. Therefore, the concentration parameter is sensitive to a \emph{higher} particle mass scale $m$.

We follow the approach of \cite{2012MNRAS.424..684S}, who parameterise the additional suppression of the concentration parameter within their WDM simulations (which includes a free streaming scale very similar to FDM) as
\begin{align}
      c(M,P_{\rm CDM})) = c^{\rm B}(P_{\rm CDM})\,\Delta_c^{\rm CDM} (M^{(c)}_0,\gamma_1,\gamma_2),
\end{align}
where $c(M,P_{\rm CDM})$ is the concentration parameter taking into account the additional suppression observed in simulations and $c^{\rm B}(M)$ is the original concentration parameter for CDM (Eq.~\ref{eq:c_of_m}) specified in \cite{bullock_2001}.
\cite{2012MNRAS.424..684S} find that the two-parameter model 
\begin{align}
    \Delta_c^{\rm CDM} (M^{(c)}_0,\gamma_1,\gamma_2)=\left[1+\gamma_1\left(\frac{f\,M^{(c)}_0}{M}\right)\right]^{-\gamma_2}
    \label{eq:corr_c_CDM}
\end{align}
provides a good fit to their simulations. 

In Eq.~\eqref{eq:corr_c_CDM} the parameter $\gamma_2$ controls the slope of the cut-off. The second parameter, $\gamma_1$, shifts the position of the cut-off scale. Note the slightly different form of the suppression function $\Delta_c^{\rm CDM}$ as compared to the HMF suppression function $\Delta_n$. 
As above, $M^{(c)}_0$ denotes the scale parameter. Note that we need to adapt a larger scale mass $M^{(c)}_0=1/f\,M^{(n)}_0$ compared to the HMF, where the parameter $f_{\rm coll}=0.01$ is used in the \cite{bullock_2001} prescription implemented in \HMC. Therefore, to be consistent with~\cite{2012MNRAS.424..684S}, we introduce an additional factor of $f$ in Eq.~\eqref{eq:corr_c_CDM}. In thin, dashed, we show the prediction using the additional suppression formula Eq.~\eqref{eq:corr_c_CDM}, with the best-fit parameters $\gamma_1=15$ and $\gamma_2=0.3$. We see that the cut-off scale is moved by a factor $\sim\gamma_1=15$ to the left compared to the scale $M^{(c)}_0$. To implement this additional suppression within \HMC, we use the following prescription
\begin{align}
      c(M,P_{\rm FDM}) =c^{\rm B}(P_{\rm FDM})\, \Delta^{\rm FDM}_c(M^{(c)}_0,\gamma_0,\gamma_1,\gamma_2)
      \label{eq:corr_c_ULA}
\end{align}
with
\begin{align}
  \Delta^{\rm FDM}_c= {\left[1+\left(\frac{M^{(c)}_0}{M}\right)\right]^{-\gamma_0}} \Delta_c^{\rm CDM}(M^{(c)}_0,\gamma_1,\gamma_2),\label{eq:Delta_c_ULA}
\end{align}
where $\gamma_0=\mathrm{d}\log c^{\rm B}(P_{\rm FDM})/\mathrm{d}\log M$, where we evaluate the derivative at $4\cdot M^{(c)}_0$, slightly before $c^{\rm B}(P_{\rm CDM})$ and $c^{\rm B}(P_{\rm FDM})$ start to deviate.  The new term in brackets in Eq.~\eqref{eq:Delta_c_ULA} counteracts the flattening of the concentration curve in the \cite{bullock_2001} model for FDM, such that without the suppression term $\Delta_c^{\rm CDM}$, we would approximately recover the CDM curve, as demonstrated by the thin, green, dashed curve in Fig.~\ref{fig:cM_model}. The suppression must first be counteracted, then suppressed again, since during a Monte Carlo parameter ``scan'' \HMC\,does not have access to the CDM power spectrum and the CDM concentration parameter, but only the axion concentration $c(M)$ (note that we drop the implicit dependence on $P_\mathrm{FDM}$ for simplicity). 

The suppression of $n(M)$ and $c(M)$ in Eqs.~\eqref{eq:corr_HMF_ULA} and \eqref{eq:corr_c_ULA} is adopted to account for a number of physical factors. As discussed above, $N$-body simulations with a free streaming scale are better fit using a sharp-$k$ window function than a real space one~\citep{Schneider:2013ria}. The shape including the sharp-$k$ function can be minimised by a real space space window with an additional suppression, while retaining the advantage of a well defined halo mass~\citep{Schive:2015kza}. \HMC\,has access to the linear theory power spectrum from \acamb for FDM, and so the suppression function must be modified to be with respect to $n^{\rm ST}(P_{\rm FDM})$. Such a function can also account for the additional suppression of halo formation caused by the Jeans scale and modified collapse barrier~\citep{Marsh:2013ywa,XiaolongThesis}. To account for the Jeans scale, the collapse barrier, $\delta_c$ should in principle also be modified, where pressure increases the overdensity required for spherical collapse~\citep[as in the case of WDM velocities as shown in ][]{2013MNRAS.428.1774B}. A modified barrier for FDM was proposed, and implemented in an approximate manner, in \cite{Marsh:2013ywa,Marsh:2016vgj}, while the excursion set for the modified barrier was solved by \cite{Du:2016aik,XiaolongThesis}. A modified collapse barrier leads to additional suppression of the mass function with respect to CDM, and also with respect to $N$-body simulations with a free streaming scale. The halo mass function has not been measured in simulations including the ``quantum pressure'' on all scales, and is likely beyond the realm of present computational ability, although see e.g. \cite{schive2014cosmic,Nori:2018pka,Li:2018kyk,Mina:2020eik,Veltmaat:2016rxo,Mocz:2019pyf}. Thus we adopt the suppression function in Eq.~\eqref{eq:corr_HMF_ULA} with the additional uncertainty shown in Fig.~\ref{fig:nM_model}. On the other hand, \cite{Marsh:2016vgj} found that the modified barrier had little effect on the non-linear power spectrum on observationally relevant scales.

In addition to these physical reasons to allow for variation in the fitted cut-offs for $n(M)$ and $c(M)$, there is also unaccounted for possible dependence on cosmological parameters, and on the method for removing spurious haloes. The required simulations to account for variability in $n(M)$ and $c(M)$ below the cut-off for ULAs are not available in the literature, and performing such simulations is beyond the scope of the present work (and is likely beyond present computational ability to resolve the required large and small scales simultaneously). We thus adopt the following wide, flat, priors on the nuisance fitting parameters:
\begin{align}
    \alpha_1 :\, &[0.6 , 2.0]\label{eqn:non-linear_nuisance} \\\nonumber
\alpha_2 : \,&[1.43 , 2.54]\\\nonumber
\gamma_1 : \,&[5.0 , 45.0]\\\nonumber
\gamma_2 : \,&[-0.37, -0.23] 
\end{align}
We illustrate the variation induced in the FDM model sampling the nuisance parameters from our priors in Figs.~\ref{fig:nM_model} and~\ref{fig:cM_model} by the green, shaded areas around  our fiducial suppression models described by Eqs.~\eqref{eq:corr_HMF_ULA} and \eqref{eq:corr_c_ULA}.  Using this method allows us in the following to present rigorous, Bayesian limits on the axion mass from weak lensing, marginalising over theoretical uncertainty in the underlying halo model.

In modelling the suppression of the HMF and the concentration mass relation, we assume that the correction functions $\Delta^{\rm FDM}_n$ and $\Delta^{\rm FDM}_c$, do not change with redshift. The fitting functions on which we base our work from \cite{Schive:2015kza} and \cite{2012MNRAS.424..684S} are reported at $z=0$ and argued to be redshift independent.

For the HMF, the cut-off at large masses, when the asymptotic value of $\sigma>1$, can be seen to be redshift independent by appeal to the known redshift independence of the $P(k)$ cut-off~\citep{Hu:2000ke}, and origin of the cut-off in the sharp-$k$ filtering model. We also compared our HMF cut-off to the fit reported by \cite{Corasaniti:2016epp}: this shows very mild redshift evolution within the uncertainties of the methods employed and accounted for by different approaches to remove spurious halos. There is also redshift dependence in the high mass end of the HMF: this is accounted for entirely in our model by using the exact $P(k,z)$ for FDM from \textsc{axionCAMB}.

For $c(M,z)$ we investigated possible redshift dependence using the alternative concentration calculation presented by \cite{Ludlow:2016ifl}, calibrated to simulations of warm DM. The method of \cite{Ludlow:2016ifl} correctly reproduces the cut-off in $c(M,z)$ seen in simulations. We applied this calculation to our FDM power spectra and observed no redshift dependence in the location of the cut-off. In future, it would be highly desirable to implement the method of \cite{Ludlow:2016ifl} into \textsc{HMCode}, but this is beyond the scope of the present work, and we find the fits of \cite{2012MNRAS.424..684S}, with the adopted uncertainty, to be sufficient, and to have little effect on our constraints.

\subsubsection{One-to-Two Halo Term Smoothing}\label{sec:smoothing}

%%%%%%%%%%%%%%%%%%%%%%%
\begin{figure}
    \centering
    \includegraphics[width=.5\textwidth]{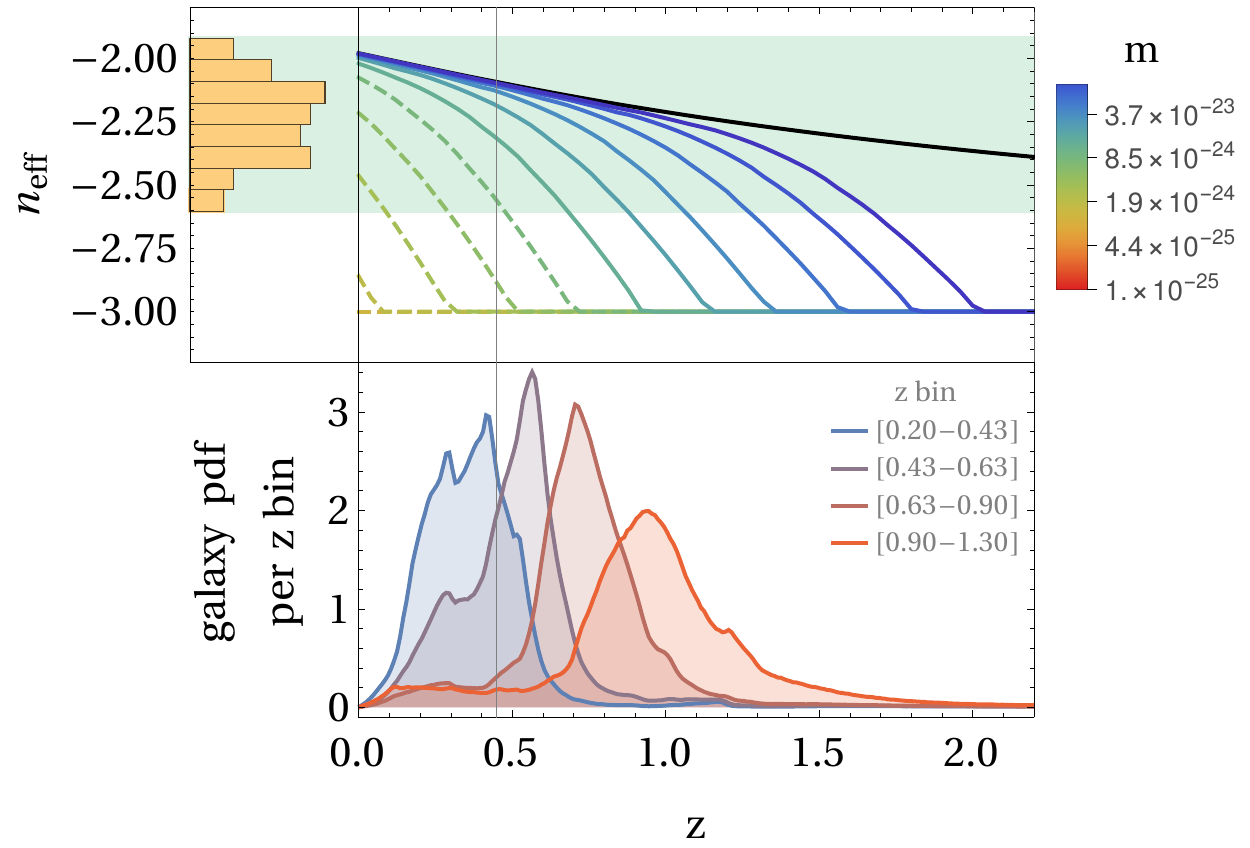}
    \caption{Evolution of $n_{\rm eff}$ as function of redshift $z$ as compared to the CDM case, which is calibrated to simulations as described in \protect\cite{Mead_2015,Mead:2020vgs}. On the far left we show a histogram of the values of $n_{\rm eff}$ determined on the nodes of the calibration emulator. In the top right panel, we show in black the evolution of $n_{\rm eff}$ in the case of CDM for  our fiducial cosmology. Colour-coded by the mass $m_{\rm Ax}$ we show the evolution of $n_{\rm eff}$ for the ULA case. In dashed we show models which are strongly excluded by our analysis. In solid we show models which are allowed or marginally excluded by our analysis. In the bottom panel, we show $n^i_{\rm src}$, the normalised redshift distribution of the numbers of galaxies in four redshift bins $i=1\dots4$. We expect the bins to be affected by capping over the redshifts that contain the most source galaxies. This implies that while the parameter $n_{\rm eff}$ leaves the calibrated range (shown by the green band in the top panels) at some $z$ for all our ULA models, this does not necessarily impact the model prediction for all the bins. As a guide to the eye, we mark the redshift at which a scale of $1\,\rm Mpc$ near the quasi-linear regime is smaller than the smallest measured angular bin at $2.8'$ for our fiducial cosmology.}
        \label{fig:effect_of_capping1}
\end{figure}
%%%%%%%%%%%%%%%%%%%%%%%%%%

\HMC\,adopts a smoothing of the transition between the one halo term and the two halo term (Eq.~\ref{eqn:one-to-two_smooth}), which captures aspects of clustering in the quasi-linear regime. The presence of this smoothing term leads to FDM displaying \emph{enhanced} power, and thus shear correlation, relative to CDM over a narrow range of masses near $m\approx 10^{-24}\text{ eV}$ (see Fig.~\ref{fig:shear_plusmminus_models_data}, and also in the the Supplementary Material, Fig.~\ref{fig:extra_models_data}). This behaviour is counter-intuitive from the point of view of the linear theory, where FDM only suppresses power.  An enhancement of power in the quasi linear regime caused by models with suppressed linear power on small scales can, however, be explained by appeal to the bias. Models with suppressed linear power generically have enhanced bias on scales above the suppression scale \citep[see e.g.][]{Carucci:2015bra,Bauer2021IntensityMapping,Lague:2021frh} due to conservation of mass. On intermediate scales, this enhancement of the bias can compensate for the linear theory suppression, leading to an overall enhancement of the power. It is possible that \HMC\,indirectly accounts for this effect via the one-to-two halo smoothing dependence on $n_{\rm eff}$ (see Eq.~\ref{eqn:neff_definition} and Table~\ref{tab:mead2015}).

However, since the \HMC\,treatment is entirely phenomenological, we must check to what extent the effect is properly calibrated for the FDM model, and to what extent an erroneous or extrapolated treatment might influence our reults. We investigate this in the following. 

The underlying parameter, $n_{\rm eff}$, that controls the one-to-two halo smoothing is calibrated to cosmological emulator nodes, where the power spectrum is computed directly from $N$-body simulations. Varying the cosmological parameters in these $N$-body simulations, can lead to a range of values for the smoothing. This is illustrated by the yellow histogram in Fig.~\ref{fig:effect_of_capping1}. The lowest value of $n_{\rm eff}$ simulated in the $N$-body simulations is $n_{\rm eff}\approx -2.6$. Fig.~\ref{fig:effect_of_capping1} also shows the value of $n_{\rm eff}(z)$ for a reference CDM cosmology (solid black curve), and for the same cosmology with various ULA masses. When the ULA linear suppression scale is larger than the non-linear scale (i.e. $k_{J,eq}<k_{nl}$), this leads to the asymptotic value $n_{\rm eff}=-3$, which is outside the range of values calibrated by simulations. We expect such models to be strongly excluded by DES, which requires the formation of non-linear cosmological structure.~\footnote{Indeed, we could even exclude such models by prior, since non-linear structure formation is required for galaxy formation and the presence of life~\citep{Weinberg:1987dv,Tegmark:2005dy}.}

Fig.~\ref{fig:effect_of_capping1} shows in the bottom panel the normalised distribution of source galaxies for DES-Y1 as a function of redshift. The grey line indicates the redshift $z\approx 0.5$ at which a scale of $1\,\rm Mpc$ is less than $2.8',$ the smallest angular bin in the DES-Y1 data.  The value of $n_{\rm eff}(z)$ computed from the FDM linear power spectrum remains fully within the calibrated range for DES source galaxies with redshifts $z>0.5$ for $m\gtrsim 10^{-23}\text{ eV}$, which, as we will see, covers all of the 95\% confidence region of our constraints. Furthermore, the parameter $n_{\rm eff}$ impacts the quasi-linear scale. For higher redshifts, this scale cannot be resolved within the range of measured angular sizes. For higher redshifts, the effect of $n_{\rm eff}$ on the model of the observables is expected to be smaller, because more of the linear part of the power spectrum and less of the quasi-linear contribute to the observables.

\begin{figure*}
    \centering
    \includegraphics[width=.85\textwidth]{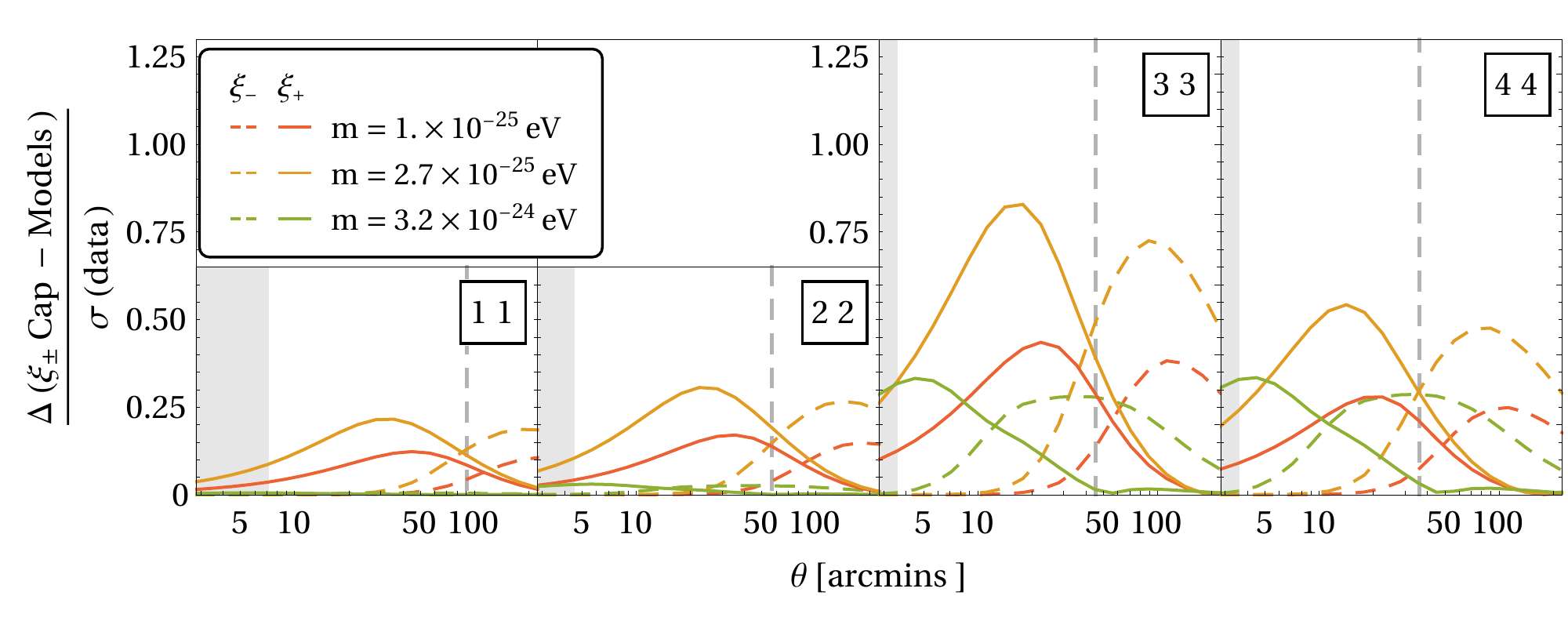}
    \caption{
    Discrepancy in the model predictions for $\xi_+$ (solid lines) and $\xi_-$ (dashed lines) for different treatments of the $n_{\rm eff}$ parameter, relative to the error (square-root of the diagonal entries of the covariance matrix). We show the diagonal bins for three different axion masses below our limit. We compare the case where we use $n_{\rm eff}$ as originally implemented in \HMC, which will leave the CDM uncalibrated region for some $z$ (see. Fig. \ref{fig:effect_of_capping1} and text for details) and the case when $n_{\rm eff}$ is capped to remain within the CDM calibrated region.  The grey band (grey, dashed line) shows $\xi_+$ ($\xi_-$) data points that are excluded from the analysis by the small scale cut imposed by the DES collaboration. For axion masses at and reasonably below our bounds, represented by the green curve, the effect of capping versus not capping is small relative to the error in the data. For $m$ far smaller than our lower limit, the effect becomes stronger, with the biggest impact on the 3-3 bin. However, these masses are in any case excluded.}\label{fig:effect_of_capping2}
\end{figure*}

To further investigate the effect on cosmic shear observables of $n_{\rm eff}(z)$ leaving the calibrated range, we computed the effect on the shear correlation function $\xi^+$ in the $i$-$i$-bins for two possibilities for $n_{\rm eff}$ (and thus smoothing parameter $\alpha$), shown in Fig.~\ref{fig:effect_of_capping2}. The first takes $\alpha$ at its value specified by $n_{\rm eff}$ according to the fits in \HMC. The second possibility we investigate applies a hard cap to $\alpha$, forcing it to lie in the range covered by the emulator nodes. We show the relative difference between the two treatments compared to the data error for an axion mass of $\sim3\times10^{-24}\,$eV (this mass is strongly excluded by our analysis, while the enhanced shear correlation caused by the smoothing parameter is maximal). As illustrated by Fig.~\ref{fig:effect_of_capping2}, the difference between the two approaches (capping or not capping the smoothing parameter) is small compared to the uncertainty in the data. 

For higher values of $m$, the difference between capping or not capping becomes smaller, and for $m>10^{-24}\text{ eV}$ is less than about 10\% the size of the DES error. Since our lower limit to $m$ will turn out to be an order of magnitude larger than this, we conclude that the effect of $n_{\rm eff}$ leaving the calibrated range of \HMC\,on our results is likely small, and will not affect our conclusions. Nonetheless in the analysis in Section~\ref{sec:dm_constraints}, we consider three possibilities for $n_{\rm eff}$: 1) leaving $n_{\rm eff}$ \emph{free}, as implemented in \HMC, 2) \emph{capping} $n_{\rm eff}$ to lie in the calibrated region, and 3) interpolating between those two possibilities by imposing an additional parameter which we \emph{marginalise} over. 

We can thus conclude that, for the purposes of limiting the FDM mass while it is fixed to be all of the dark matter, we can safely take the prescription for one-to-two halo smoothing as specified by \HMC, since the power spectrum slope falls within the range of the cosmologies covered by the emulator nodes for FDM masses within the bulk of the posterior probability, and the effect of capping the smoothing parameter has an effect on the correlation function smaller than the observational error bars. Nonetheless, the calibration of \HMC\,for beyond CDM models in the quasi-linear regime is an important open problem and should be investigated further using simulations. It will be particularly important to search for evidence of sub-dominant ULAs with $m<10^{-23}\text{ eV}$ using high precision lensing observables from e.g. \emph{Euclid}~\citep{Marsh:2011bf,Amendola:2016saw}.

Finally, we note that the smoothing term adopted in \HMC\,is not present in the comparably accurate halo model/perturbation theory hybrid of \cite{Sullivan:2021sof}. The hybrid model does, however, contain elements that might account for the same features in $P(k)$ captured by the \HMC\,smoothing term. In the hybrid model the two halo term is modified to follow the Zel'dovich approximation (Lagrangian perturbation theory), while the one halo term is modified to include a ``broadband expansion'' of the halo profile, which has free coefficients to be fitted to simulations, and can be used to model baryonic feedback. The hyrbid model can account for enhancement of power in the quasi-linear regime (as caused by \HMC\,smoothing when $n_{\rm eff}$ is small)  via a polynomial transfer function with further fitting parameters augmenting the Zel'dovich approximation. The coefficients of all these fits to simulations are found to vary with cosmological parameters, but \cite{Sullivan:2021sof} do not vary $\sum m_\nu$ or $n_s$, both of which would affect $n_{\rm eff}$ dependence of the fitting parameters in the transition region. Thus it is not possible to assess whether the hybrid model would similarly predict the observed effects of FDM in \HMC\,based on the fits presented thus far by \cite{Sullivan:2021sof}. It would be useful in further application of the hybrid model to investigate parameter dependences of the fitting based on model independent phenomenological parameters such as $n_{\rm eff}$, which can be applied to any input linear power spectrum.

%%%%%%%%%%%%%%%%%%%

%%%%%%%%%%%%%%%%%%%%%%%%%%%%%%%%%%%%%%%%%%%%%%%%%%%%%%%%%%%%%%%%%%%%%%%%%%%%%%%%%
\section{Data and Methodology}
\label{sec:data}
%%%%%%%%%%%%%%%%%%%%%%%%%%%%%%%%%%%%%%%%%%%%%%%%%%%%%%%%%%%%%%%%%%%%%%%%%%%%%%%%%

\subsection{Statistical Methods}
\label{sec:stats_methods}

We perform Bayesian analysis using the publicly available code \cosmosis to implement the DES-Y1 and \emph{Planck} likelihoods. We use \textsc{multinest}~\citep{Feroz:2008xx} as implemented within \cosmosis to derive parameter constraints. We used a stopping criterion of $\log{Z}=10^{-5}$ and 2100 (3500 for the full ``3x2pt" analysis, see the Supplementary Material) livepoints to ensure smooth confidence intervals from the \textsc{multinest} chains.

On a subset of runs, we also compared the \textsc{multinest} constraints with Markov Chain Monte Carlo (MCMC) analysis using the affine invariant \emcee~\citep{ForemanMackey:2012ig} sampler as a cross-check. We used these tests to also perform convergence tests of our result using the spectral method~\citep{Dunkley:2004sv}, and to compute the autocorrelation time of our chains. 

\begin{figure}
    \centering
    \includegraphics[width=0.8\linewidth]{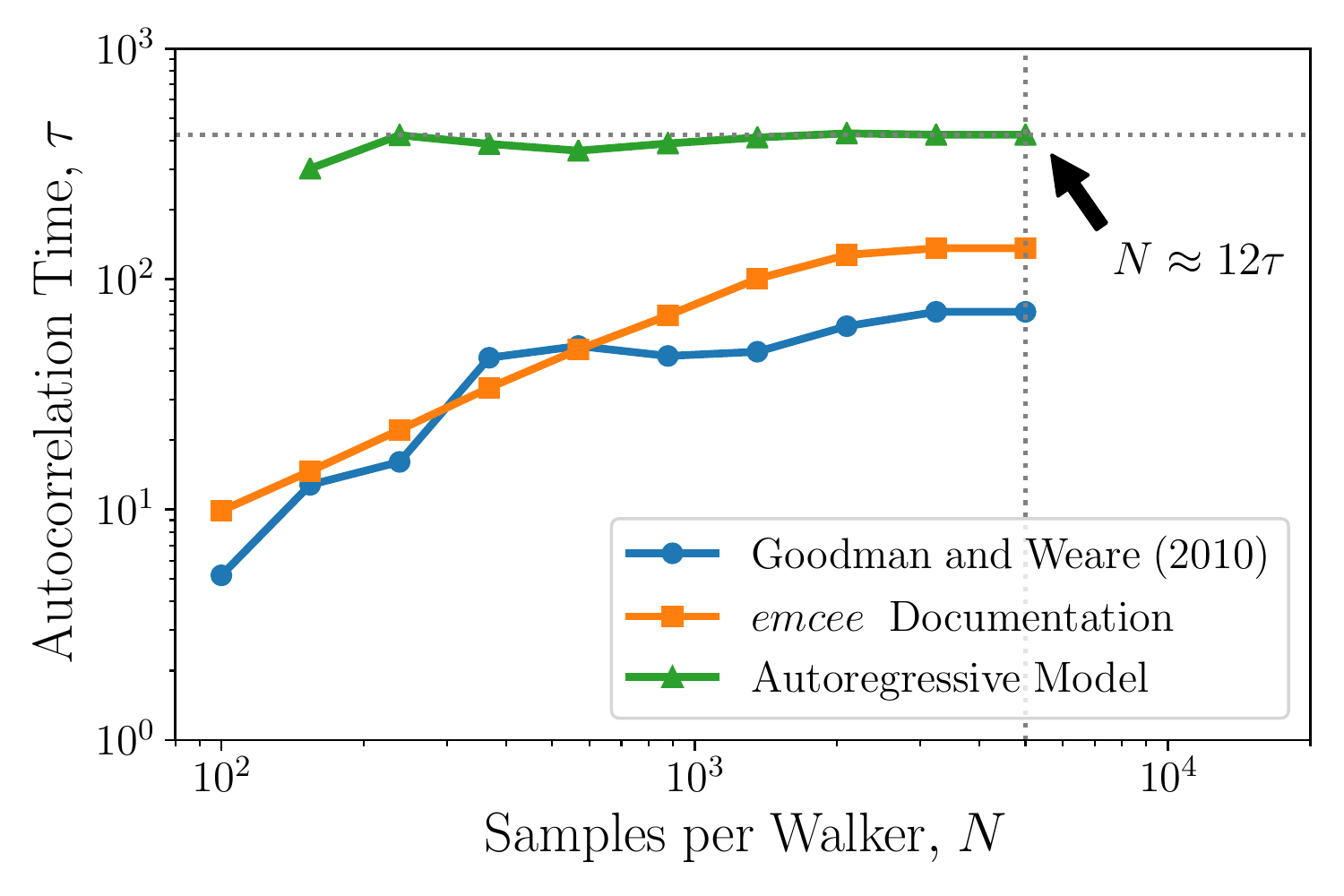}
    \caption{Estimation methods for the chain autocorrelation times. We show that the autoregressive model approach gives a more stable estimate of the autocorrelation time with short chains. The fact that the autoregressive model slightly overestimates the autocorrelation time means that this method to evaluate convergence is conservative since more samples are needed to reach $N\gg \tau$.}
    \label{fig:autocorr_analysis}
\end{figure}

The autocorrelation method is useful for chains with multiple walkers, because the error on the MCMC estimate for a given parameter is proportional to $\tau/N$ where $N$ is the total number of samples (i.e. the number of samples per walker times the number of walkers) and where $\tau$ is the integrated autocorrelation time~\citep{Goodman2010:EnsembleSamplers}. We found that the standard autocorrelation time convergence criterion described in the \emcee documentation to be overly stringent for our cases with a large number of parameters and walkers, and time consuming likelihood and theory calculations. Specifically, finding a reliable estimate of the autocorrelation time would have required a prohibitively large number of steps and computation time. We thus used to an autoregressive model to get an estimate of the autocorrelation time with fewer samples. After verifying that the autoregressive model did not underestimate the autocorrelation time, we ensured that all the chains satisfied $N \gg\tau$ using that estimate. An example comparison of the autocorrelation time estimates from the three methods described above can be found in Fig.~\ref{fig:autocorr_analysis}.

\subsection{DES Data Sets}
Our analysis is based on the DES-Y1 3x2pt data set consisting of the observables $\xi_\pm$, $\gamma_t$ and $w$ described in Section (\ref{sec:DES_observables}). Consistent with the discussion there, we focus on the cosmic shear data in our main analysis,  because it does not depend on the galaxy bias and is therefore less sensitive to modelling uncertainties.
The DES data on the cosmic shear observables in the $4,4$ bin was shown already in Fig.~\ref{fig:shear_powerspectra} illustrating the effect of the FDM mass holding all other parameters fixed. For the respective figures the other bin combinations as well as the figures of $\gamma_t$ and $w$ we refer to the Supplementary Material.

We validated our pipeline for the DES-Y1 data with a series of cross checks, described briefly here. First, we investigate the effects of our non-linear model on CDM by comparing our constraints on the cosmological parameters using \HMC\,and the alternative non-linear model \textsc{Halofit}~\citep{2003MNRAS.341.1311S}, using both 3x2pt and shear only, finding no significant differences. As noted in \cite{Hlozek:2016lzm}, \textsc{Halofit} cannot be applied to ULAs/FDM. Next, we compared the results of our CDM analysis to those of the publicly available chains from  DES~\footnote{\url{http://desdr-server.ncsa.illinois.edu/despublic/y1a1_files/chains/}}, finding visual agreement in the two-dimensional posteriors. 

As discussed by \cite{Joudaki:2019pmv}, the marginalised limits on the main parameters constrained by weak lensing, ($\sigma_8$,$\Omega_m$), are prior dependent: the upper limit on $\sigma_8$ depends on how $A_s$ is sampled, and on the assumptions made about neutrino mass. Furthermore, the limit on $\Omega_m$ relies on the 3x2pt data combination, and thus depends on the galaxy bias model. Using weak lensing as an anchor to the non-linear scales to constrain DM thus requires further inputs for the cosmological parameters.

\subsection{Planck Data and Data Combination}

The full \emph{Planck} data set of two point statistics includes temperature $T$, $E$ and $B$ mode polarisation, and lensing deflection $\phi$ power spectra and cross-spectra. The \emph{Planck} lensing deflection is correlated with DES lensing observables. We do not model this correlation, and so we omit $\phi$ auto and cross spectra from our \emph{Planck} data. Small angle polarisation $B$ modes are generated primarily from lensing, and so we omit these also from our analysis. We neglect the covariance between the \emph{Planck} \(TT\) angular power spectrum and DES weak lensing arising from the lensing-induced smoothing of small-scale \(TT\) peaks and troughs. We consider only the adiabatic mode of initial conditions, as described in \cite{Hlozek:2014lca}, omitting isocurvature modes. A complete analysis of \emph{Planck} data in the ULA model including lensing and isocurvature was performed by \cite{Hlozek:2017zzf}. For the particle masses of interest, omitting isocurvature modes is equivalent to a prior on the Hubble scale during inflation $H_I\lesssim 10^{12}\text{ GeV}$ \citep{Marsh:2015xka,Hlozek:2017zzf}.

In our analysis we combine the \emph{Planck} CMB data with DES-Y1 for both shear only and 3x2pt. In this combination, the \emph{Planck} CMB data constrains the six standard cosmological parameters extremely well, and for CDM the effect of DES-Y1 is only to slightly tighten the error bars by relatively small amounts. Applied to constraining the nature of DM, however, the utility of the combination is greater: \emph{Planck} anchors the large scales and the cosmological parameters, while DES anchors smaller scales and thus constrains the nature of DM further. 
\begin{table}
    \centering
    \begin{tabular}{c|c}
         Parameter& Prior range  \\
         \hline 
         DM particle mass $\log{m}/\mathrm{eV}$&$\mathcal{U}[-25.5, -22]$ \\
         DM density $\Omega_{d}$&$\mathcal{U}[0.1, 0.9]$\\
         Hubble parameter $H_0$ [km/s/Mpc]&$\mathcal{U}[55, 90]$\\
         Baryon density $\Omega_b$&$\mathcal{U}[0.03, 0.07]$\\
         Scalar spectral index $n_s$ & $\mathcal{U}[0.87, 1.07]$\\
         Scalar amplitude $A_s$& $\mathcal{U}[5e^{-10}, 5e^{-9}]$\\
         Neutrino density $\Omega_\nu h^2$&$\mathcal{U}[0, 0.01]$ \\
         \hline
    \end{tabular}
    \caption{Cosmological parameters varied in the analysis.}
    \label{tab:cosmo}
\end{table}

%%%%%%%%%%%%%%%%%%%%%%%%%%%%%%%%%%%%%%%%%%%%%%%%%%%%%
\subsection{Parameters, Priors, and Models}
\label{sec:statistics}

We adopt primary cosmological parameters:
\begin{equation}
    \{A_s,n_s,H_0, \Omega_b,\Omega_m,\sum m_\nu \}\, , \label{eqn:cosmo_params}
\end{equation}
where $A_s,n_s$ are the primordial power spectrum amplitude and spectral index, $H_0$ is the present day Hubble parameter, $\Omega_b$ is the baryon density, and $\Omega_m$ is the total matter density, $\Omega_m=\Omega_b+\Omega_d$ and the DM ($\Omega_d$) is composed of either CDM (for test cases) or FDM, but \emph{not} both. Massive neutrinos are included with the parameter $\sum m_\nu$ for which we assume a single massive neutrino and $N_{\rm eff}=2.04$ massless neutrinos. The neutrino mass is related to the neutrino density parameter by $\Omega_\nu h^2=\sum m_\nu/94.\text{ eV}$. For detailed discussion of how massive neutrinos are treated in the halo model, we refer to \cite{Mead_2015,Mead:2020vgs}.

The parameters are varied with flat priors as specified in Table~\ref{tab:cosmo}. These priors are consistent with the choices made by the DES collaboration in their own analysis. These are different from the priors adopted in standard CMB analyses, which use a log prior on $A_s$, vary the physical densities $\Omega_c h^2$ and $\Omega_b h^2$, and $H_0$ is a derived parameter from the angular scale of the first acoustic peak. The parameter choices of Eq.~\eqref{eqn:cosmo_params} are not optimal for a CMB analysis. When analysing galaxy data only, \cite{Joudaki:2019pmv} showed how these different priors affect the resulting bounds on $S_8$ and $\Omega_m$. When we combine DES-Y1 with \emph{Planck}, the prior dependence largely vanishes due to the strong constraining power of \emph{Planck} for the power spectrum amplitude and matter density. When using the \emph{Planck} data, we vary the optical depth $\tau$ with a uniform prior, and the \emph{Planck} absolute calibration parameter, $A_{\rm Planck}$, with a Gaussian prior.

The DES-Y1 analysis contains a large number of nuisance parameters, which are listed in Table~\ref{tab:des_nuisance}. These are varied, and marginalised over in any data combination including DES.

\begin{table}
    \centering
    \begin{tabular}{c|c|c}
         Parameter& Sampling range & Prior  \\
         \hline
         \multicolumn{3}{c}{Bin Bias parameters}\\
         $b_1$   &$\mathcal{U}[0.8, 3.0]$& $\mathcal{U}[0.8, 3.0]$\\ 
         $b_2$& $\mathcal{U}[0.8, 3.0]$& $\mathcal{U}[0.8, 3.0]$\\
         $b_3$ &$\mathcal{U}[0.8, 3.0]$ &$\mathcal{U}[0.8, 3.0]$\\
         $b_4$ & $\mathcal{U}[0.8, 3.0]$& $\mathcal{U}[0.8, 3.0]$\\
         $b_5$ & $\mathcal{U}[0.8, 3.0]$& $\mathcal{N}[0.8, 3.0]$\\
           \hline
          \multicolumn{3}{c}{Shear calibration parameters}\\
        $m_1$ & $\mathcal{U}[-0.1, 1]$&$\mathcal{N}[0.012, 0.023]$\\
        $m_2$ & $\mathcal{U}[-0.1, 1]$&$\mathcal{N}[0.012, 0.023]$\\
        $m_3$ & $\mathcal{U}[-0.1, 1]$&$\mathcal{N}[0.012, 0.023]$\\
        $m_4$ & $\mathcal{U}[-0.1, 1]$&$\mathcal{N}[0.012, 0.023]$\\
            \hline
         \multicolumn{3}{c}{Intrinsic alignment parameters parameters}\\
         $a$ & $\mathcal{U}[-5, 5]$& $\mathcal{U}[-5, 5]$\\
         $\alpha$ &$\mathcal{U}[-10, 10]$ &$\mathcal{U}[-10, 10]$\\
             \hline
          \multicolumn{3}{c}{Weak lensing photo-$z$ error parameters}  \\
          $wl-pz_1$& $\mathcal{U}[-0.1, 1]$& $\mathcal{N}[-0.001, 0.016]$\\
          $wl-pz_2$& $\mathcal{U}[-0.1, 1]$& $\mathcal{N}[-0.019, 0.013]$\\
          $wl-pz_3$ & $\mathcal{U}[-0.1, 1]$& $\mathcal{N}[0.009, 0.011]$\\
          $wl-pz_4$& $\mathcal{U}[-0.1, 1]$& $\mathcal{N}[-0.018, 0.022]$\\
              \hline
        \multicolumn{3}{c}{Lensing photo-$z$ error parameters}  \\
    $l-pz_1$ & $\mathcal{U}[-0.05,0.05]$& $\mathcal{N}[0.008, 0.007]$\\
    $l-pz_2$ & $\mathcal{U}[-0.05,0.05]$& $\mathcal{N}[-0.005, 0.007]$\\
    $l-pz_3$ & $\mathcal{U}[-0.05,0.05]$& $\mathcal{N}[0.006, 0.006]$\\
    $l-pz_4$& $\mathcal{U}[-0.05,0.05]$& $\mathcal{N}[0.00, 0.01]$\\
    $l-pz_5$&$\mathcal{U}[-0.05,0.05]$& $\mathcal{N}[0.0, 0.01]$ \\
     \hline
        \multicolumn{3}{c}{CMB Parameters}  \\
        $\tau$&$\mathcal{U}[0.02,0.2]$& $\mathcal{U}[0.02,0.2]$\\
        $A_{\rm Planck}$&$\mathcal{U}[0.9,1.1]$& $\mathcal{N}[1, 0.0025]$\\
        \hline
    \end{tabular}
    \caption{The nuisance parameters for the DES-Y1 and Planck18 dataset, and their sampling priors.}
    \label{tab:des_nuisance}
\end{table}
The axion mass is varied as:
\begin{equation}
    \mathcal{P}(\log_{10}[m/\text{eV}]) = \mathcal{U}[-25.5,-22]\, . \label{eqn:log-flat_m}
\end{equation}
Such a log-flat prior, while arguably less informative than alternatives, must have upper and lower limits to keep the prior volume finite. The lower bound in Eq.~\eqref{eqn:log-flat_m} is motivated by the CMB constraints varying the axion fraction \citep{Hlozek:2017zzf}, who find that at our lower limit the CMB forbids axions from being all the DM at high significance. Lower masses have vanishing posterior weight, and our exclusion limit is not affected by the lower limit of the prior extending to smaller values. Since we expect a one sided lower limit to $\log_{10}(m/\text{eV})$, however, the upper bound of the prior will affect the exclusion. In such a case, the upper bound of the prior should be chosen to be close to the expected experimental sensitivity, which can in principle be computed \emph{a priori} knowing only the experimental specifications (e.g. from a Fisher matrix forecast). 

We set the upper limit of our prior following the discussion in the introduction, as follows. We wish to set our prior upper limit to be equivalent to CDM, but to establish in a data-driven manner. \cite{Chabanier:2019eai} project the DES-Y1 data into wavenumber bins on the linear power spectrum, the locations of which can be computed without specifying the measurement value in the bin. The highest bin for DES-Y1 covers $1\, h\,\text{Mpc}^{-1}<k<5\, h\,\text{Mpc}^{-1}$. Setting $k_{\rm J,eq}(m)=5 \, h\,\text{Mpc}^{-1}$ in Eq.~\eqref{eqn:kj_eq} gives $m=1.5\times 10^{-23}\text{ eV}$ as the approximate DES sensitivity to $m$. Due to the rough nature of this estimate, we choose our maximum axion mass to be an order of magnitude larger than the expected sensitivity. A log-flat prior on $m$ was also used in \cite{Rogers:2020ltq}, where the lower limit was also set according to existing bounds, and the upper limit was set according to the smallest scale (i.e. heaviest axion mass) to which the Lyman-$\alpha$ forest data were sensitive.
\begin{figure*}
    \centering
    \includegraphics[width=.7\textwidth]{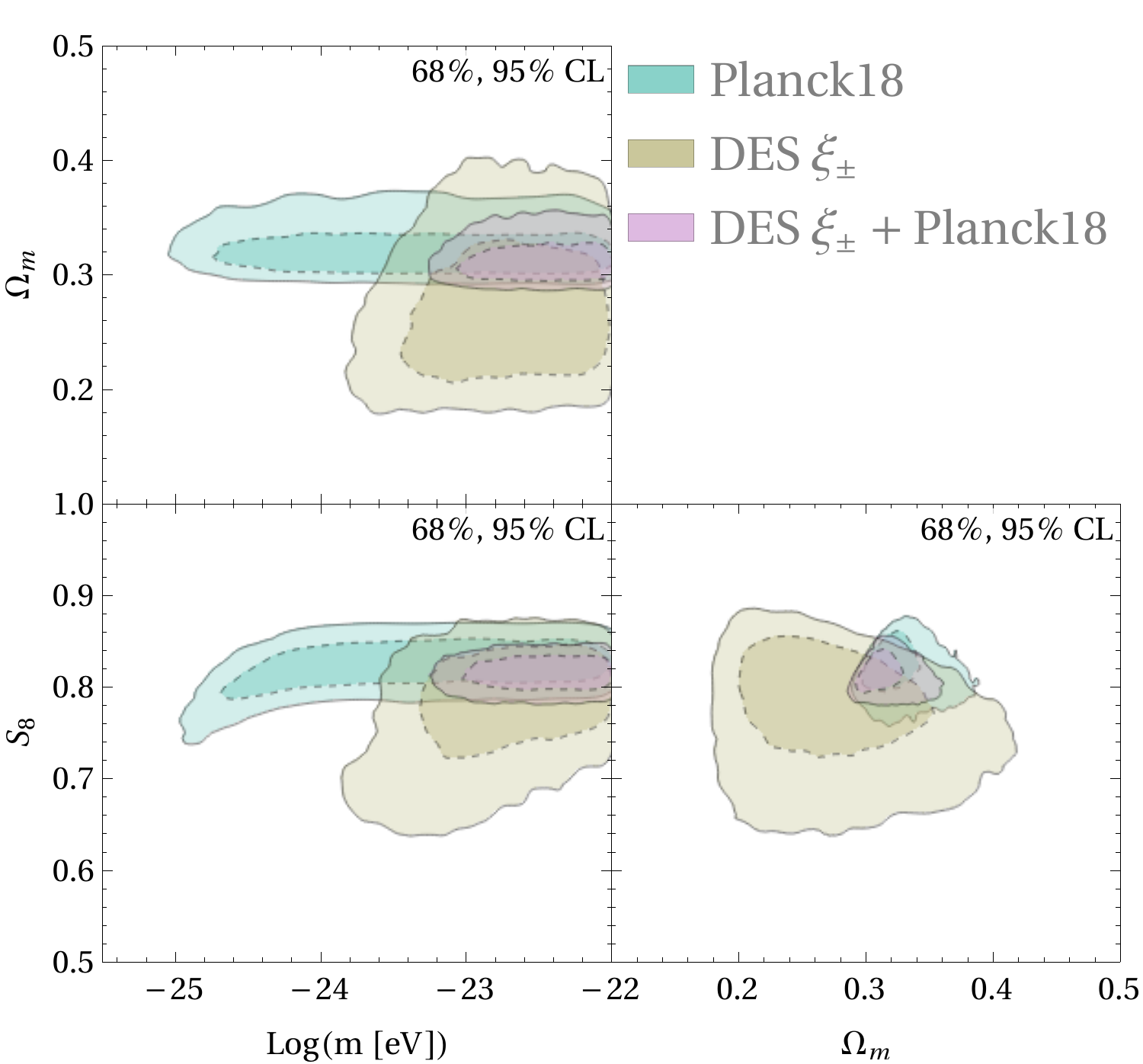}
    \caption{Posterior density for the DM particle mass $m$, the mass fluctuation parameter $S_8$, and the matter density $\Omega_m$ for different combinations of data sets.}
    \label{fig:axion_cosmo}
\end{figure*}
The disadvantage of a log-flat prior is that there is infinite prior volume between any chosen upper limit and the CDM limit $m\rightarrow \infty$. The alternative prior:
\begin{equation}
    \mathcal{P}(10^{-23}\text{ eV}/m) = \mathcal{U}[0,200]\, , \label{eqn:flat_invm}
\end{equation}
has finite prior volume up to the CDM limit, $\mathcal{P}(0)$. Such a prior has been considered constraining the dark matter nature from the Lyman-alpha forest~\citep[e.g.][]{Irsic:2017ixq}. However, this inverse mass prior requires the choice of a reference mass scale set by the lower limit and, unlike the log-flat prior, it does not apply equal prior probability to different logarithmic particle mass scales within the limits set by existing bounds and projected sensitivity. In the following, our main analysis uses the log-flat prior Eq.~\eqref{eqn:log-flat_m}, but we also present a comparison to the case Eq.~\eqref{eqn:flat_invm} for a subset of our analyses.

We adopt various nuisance parameters to account for uncertainty in the non-linear model of the HMF and halo concentration, as given in Eq.~\eqref{eqn:non-linear_nuisance}. In a subset of our analysis we also investigated a nuisance parameter for the halo model smoothing parameter, $\alpha$, as described in Section~\ref{sec:smoothing}.

%%%%%%%%%%%%%%%%%%%%%%%%%%%%%%%%%%%%%%%%%%%%%%%%%%%%%%%%%%%%%%%%%%%%%%%%%%%%%%%%%
\section{Results}\label{sec:dm_constraints}
%%%%%%%%%%%%%%%%%%%%%%%%%%%%%%%%%%%%%%%%%%%%%%%%%%%%%%%%%%%%%%%%%%%%%%%%%%%%%%%%%
%%%%%%%%%%%%%%%%%%%%%%%%%%%%%%%%%%%%%%%%%%%%%%%%%%%%%%%%%%%%%%%%%%%%%%%%%%%%%%%%%

Our baseline analysis uses the combination of \emph{Planck} $TT$, $TE$, and $EE$ spectra and DES-Y1 shear. Fig.~\ref{fig:axion_cosmo} shows marginalised two dimensional posteriors on the FDM mass and cosmological parameters for our baseline analysis. We show the total matter density $\Omega_m=\Omega_d+\Omega_b$, and the ``matter fluctuation parameter'' $S_8=\sigma_8\,(\Omega_m/0.3)^{0.5}$. These are the parameters of most interest for weak lensing surveys~\citep{Abbott:2017wau}, and the other parameters are well constrained in combination with \emph{Planck}.

For \emph{Planck} alone, there is a slight degeneracy between $S_8$ and $m$ at very low masses, due to the suppression of the linear power caused by FDM. Fig.~\ref{fig:axion_cosmo} demonstrates that, due to the tight constraints on the primary parameters provided by the \emph{Planck} data, there is no degeneracy between the ULA mass and other cosmological parameters at higher mass (the same is true of the cosmological parameters that are not shown in this figure). For DES alone there are stronger degeneracies, and wider limits to $S_8$ and $\Omega_m$. 
 
Compared to \emph{Planck} alone, adding in the DES-Y1 data provides an anchor to measure the power spectrum into the non-linear regime, leading to an improved lower bound to the FDM mass:
\begin{align}
    \log_{10} (m/\text{eV})&\geq -23.0 \quad \text{(95\% C.L.)}\nonumber \\
     \Rightarrow m &\geq 1\times 10^{-23}\text{ eV}\, .
\end{align}
The marginalized posterior distributions for the axion mass for \textit{Planck} and DES are shown in Fig.~\ref{fig:posterior_1d}. Both probes show an agreement with $m>10^{-22}$ eV, and are therefore consistent with DM being cold in a model allowing FDM. DES alone has a peak probability at $m\approx 10^{-22.5}$ eV which we attribute to the lower values of $S_8$ and $\Omega_m$ measured by this survey in isolation. In the posterior for DES alone, we observe a peak probability at $m\approx 10^{-22.5}$ eV which we attribute to a \textsc{multinest} sampling artefact, which has no effect on the value of the exclusion limit (the limits agree between \textsc{multinest} and \textsc{emceee} and no corresponding peak is present with \textsc{emcee}).

%%%%%%%%%%%%%%%%%%%%%%%%%%%%%%%%%%%%%%%%%
% Two halo smoothing comparison plot
%%%%%%%%%%%%%%%%%%%%%%%%%%%%%%%%%%%%%%
\begin{figure*}
    \centering
    \includegraphics[width=0.27\textwidth,angle=90,  trim={7cm 0cm 7cm 0cm}, clip]{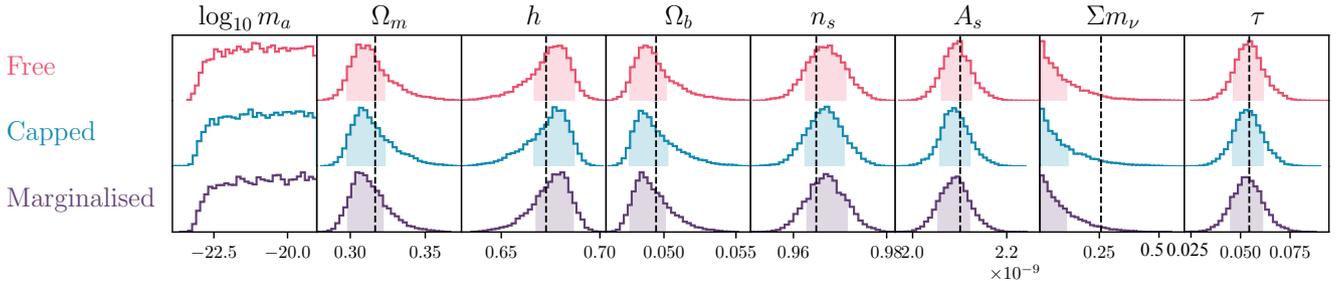}
    \caption{Cosmological parameters and axion mass posteriors with three different treatments of the two halo smoothing parameter, $n_{\rm eff}$, for \emph{Planck}+DES-Y1 shear (see text for description). Consistency between the treatements demonstrates that the additional effects caused by $n_{\rm eff}$ leaving the range calibrated to simulation do not affect our conclusions related to the cosmological parameters and FDM mass. The vertical dashed lines show the best-fit parameter values from the published Planck TT,EE,TE+lowE results \protect\cite{Planck:2018vyg}, and the 95\% upper bound for the neutrino mass. Shifts in the central values compared to \emph{Planck}, though consistent at $1\sigma$ (shaded region), are caused by our adopting priors consistent with the DES-Y1 analysis, but different from the \emph{Planck} analysis \protect\citep[see discussion in][]{Joudaki:2019pmv}.}
    \label{fig:neff_marg_compare}
\end{figure*}

As discussed in Section~\ref{sec:smoothing}, the two-halo smoothing term in \HMC\,is controlled by a parameter, $n_{\rm eff}$, which for FDM leaves the range calibrated to simulation. To estimate the dependence of our cosmological results on the smoothing term, we ran three separate analyses treating $n_{\rm eff}$ differently: one in which it was \emph{free} to take the value outside of the range calibrated through simulations and predicted by the FDM linear power spectrum, one in which $n_{\rm eff}$ was \emph{capped} to remain within the calibrated range, and one in which it was \emph{marginalised} with an additional nuisance parameter if it left the calibrated range. 

We compared the maximum likelihood in a range of mass bins for the different models for $n_{\rm eff}$. As anticipated in our discussion in sec.~\ref{sec:smoothing}, we find that the different models for $n_{\rm eff}$ impact the likelihood at small masses to a small extent. At higher masses, near to where we set our exclusion limits, the effect is negligible for the combined DES and Planck18 likelihood and very small for the DES likelihood alone. We then proceeded to check the effect of the three different treatments of $n_{\rm eff}$ on the posterior. For DES alone, we found a weak dependence on the $n_{\rm eff}$ parameter in the 95\% region of the two dimensional marginal posteriors, which is not present in the combined analysis. The one-dimensional posteriors for \emph{Planck}+DES-Y1, for all three treatments, are shown in Fig.~\ref{fig:neff_marg_compare}. These results are consistent between all three treatments, demonstrating that our conclusions are not sensitive to the treatment of the smoothing term outside the range calibrated to simulation. 
%%%%%%%%%%%%%%%%%%%%%%
\begin{figure*}
    \centering
    \includegraphics[width=0.27\textwidth,angle=90,  trim={7.0cm 0cm 7.0cm 0cm}, clip]{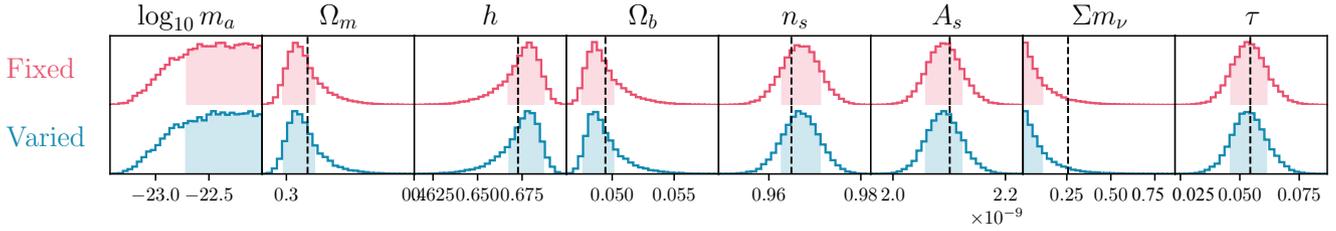}
    \caption{Cosmological parameters and axion mass posteriors with two different treatments of the halo mass function, $n(M)$, and concentration mass relation, $c(M)$, cut-offs (see Figs.~\ref{fig:cM_model} and \ref{fig:nM_model}). In the upper panels, the cut-offs are fixed to their fiducial values following simulations in \protect\cite{2012MNRAS.424..684S,Schive:2015kza}, while the lower panels additional uncertainty in the low mass shape is marginalised. Consistency between the treatments demonstrates that the additional effects caused by uncertainty in $n(M)$ and $c(M)$ below the cut-off scale do not affect our conclusions related to the cosmological parameters and FDM mass. The vertical dashed lines show best-fit values from the \textit{Planck} (TT,EE,TE+lowE) data combination \protect\cite{Planck:2018vyg}, and the 95\% confidence limit for the neutrino mass. Shifts in the central values compared to \emph{Planck}, though consistent at $1\sigma$ (shaded region), are caused by our adopting priors consistent with the DES-Y1 analysis, but different from the \emph{Planck} analysis \protect\citep[see discussion in][]{Joudaki:2019pmv}.}
    \label{fig:marg_HM_params}
\end{figure*}
%%%%%%%%%%%%%%%%%%%%%%%%%%

In addition, Fig.~\ref{fig:neff_marg_compare} also demonstrates the consistency between our results with FDM and the \emph{Planck} analysis, which assumes CDM. We also verified this independently with our own CDM analysis, finding that the marginalised posteriors on the cosmological parameters align for CDM and ULAs. This implies, qualitatively, that there is little to no effect on $\sigma_8$ and $H_0$ tensions including FDM as all the DM in the mass range we have considered.

Similarly to the uncertainty in the halo model smoothing parameter, we found that our analysis was also insensitive to the other nuisance parameters listed in Eq.~\ref{eqn:non-linear_nuisance}, which are introduced to account for uncertainty in the HMF and $c(M)$. The nuisance parameters introduced to account for this uncertainty show no degeneracy with any cosmological parameters, in particular they are not degenerate with the FDM mass. The marginalised posterior on the axion mass does not change when these parameters are varied compared to the case when they are held fixed, Fig.~\ref{fig:marg_HM_params}. 

From these runs varying the parameters controlling the uncertainty on the FDM non-linear power, we conclude that these uncertainties lead to no significant effect on the FDM mass constraint derived from DES-Y1 weak lensing measurements at the current level of precision in the data. 

%%%%%%%%%%%%%%%%%%%%%%
% Varying mass priors plot
%%%%%%%%%%%%%%%%%%%%%
\begin{figure*}
\begin{center}
\includegraphics[width=0.8\textwidth]{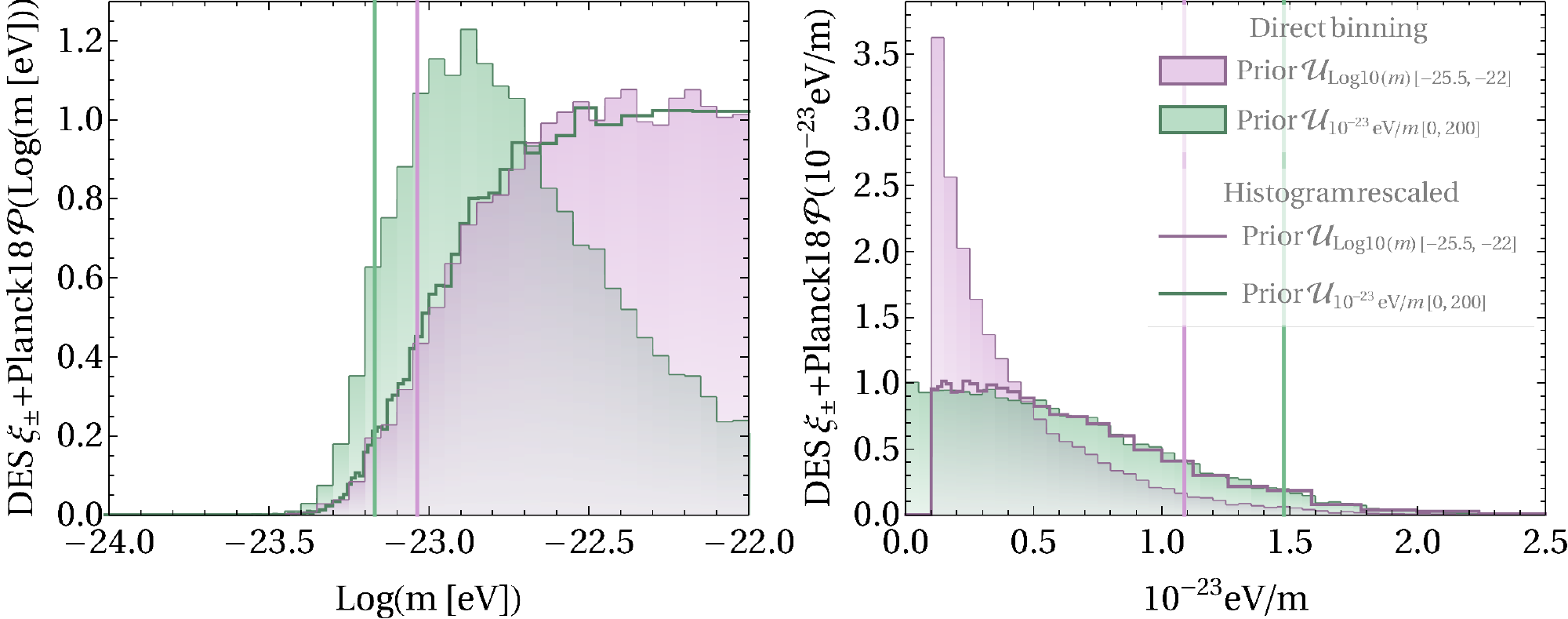}
\caption{Prior dependence. Purple histograms show the case of a log-flat prior on the particle mass, while green histograms show the case of a uniform prior on the inverse mass. The left panel shows the resulting posterior in log-space, while the right panel shows the posterior in $1/m$ space. In each case we find a lower/upper limit to the variable in the space over which the prior was uniform, but we find a different shaped posterior after a change of variables.  For each prior, we mark the 95\% limit by a vertical line to illustrate the noticeable difference in the derived mass bound. We also show as un-filled histograms (solid lines) the case of binning our chains in the original parameter and transforming the histogram bounds to the complementary parameter. After this rescaling (effectively a change of prior), these transformed histograms agree very well with the posteriors obtained from the  complementary prior. This illustrates that the different results for the mass parameter $m$ are indeed caused by the different priors and are not a sampling artefact or an effect in the likelihood.
This illustrates that care must be taken to include knowledge of the prior volume when producing mass limits after a changeof variable.}\label{fig:prior}
\end{center}
\end{figure*}
%%%%%%%%%%%%%%%%%%%%

Fig.~\ref{fig:prior} demonstrates the effect on the posterior distribution from the prior on the FDM mass (see \S~\ref{sec:statistics} for more discussion). Our \textsc{multinest} chains correspond to samples drawn from the posterior distribution of the parameters, in particular the mass parameter $m$, sampled in $\log_{10}(m/\text{eV})$ and $10^{-23}\,\text{eV}/m$, respectively. When we calculate the 95\% percentile of these distributions, we find a mass bound of $m=9.2\times10^{-24}\,$eV for the log-flat prior and $m=6.7\times10^{-24}\,$eV for the prior flat in  $1/m$ when transformed to limits in $m$ itself. We can see how this difference comes about by binning our chains evenly in both mass parameters, $\log(m)$ and $1/m$, where we convert from one mass parameter to the other \emph{before} binning the samples from our chains. The resulting histograms, illustrated by filled histograms in Fig.~\ref{fig:prior}, differ notably when a change of variables is performed. In particular, we notice a peak in the posterior around $10^{-23}\text{ eV}$ when transforming the $1/m$ prior to log space. This is caused entirely by the choice of scale inherent in the $1/m$ prior, and is not a feature of the data.

To demonstrate that this is caused by the choice of prior, rather than a feature in the data/likelihood or a sampling artefact, we furthermore evenly bin our chains in the respective original mass parameter and \emph{subsequently} convert the resulting histogram bounds to the complementary mass parameter. This process is equivalent to importance re-sampling, where we would re-weight each element of the Multinest chain by the ratio of the respective priors. For our special case of two flat priors, the respective posteriors agree up to normalization within the range of the two priors, therefore we can directly transform the histogram, which represents our estimate of the posterior. Indeed, we find that after suitable normalisation, the resulting histogram agrees well with the posterior distribution obtained from the complementary parameter. This illustrates that there are no features in the data leading to the different shapes of distribution. In a Bayesian analysis one must always present limits to the parameter which is varied and be careful comparing results using different priors. Our log-uniform prior selects a scale only in the data-driven upper limit, as discussed above, and therefore has physical motivation.

%%%%%%%%%%%%%%%%%%%%%%%%%%%%%%%%%%%%%%%%%%%%%%%%%%%%%%%%%%%%%%%%%%%%%%%%%%%%%%%%%
%%%%%%%%%%%%%%%%%%%%%%%%%%%%%%%%%%%%%%%%%%%%%%%%%%%%%%%%%%%%%%%%%%%%%%%%%%%%%%%%%
%
%
%%%%%%%%%%%%%%%%%%%%%%%%%%%%%%%%%%%%%%%%%%%%%%%%%%%%%%%%%%%%%%%%%%%%%%%%%%%%%%%%%
%%%%%%%%%%%%%%%%%%%%%%%%%%%%%%%%%%%%%%%%%%%%%%%%%%%%%%%%%%%%%%%%%%%%%%%%%%%%%%%%%
\section{Discussion and Conclusions}
\label{sec:discussion}
%%%%%%%%%%%%%%%%%%%%%%%%%%%%%%%%%%%%%%%%%%%%%%%%%%%%%%%%%%%%%%%%%%%%%%%%%%%%%%%%%
%%%%%%%%%%%%%%%%%%%%%%%%%%%%%%%%%%%%%%%%%%%%%%%%%%%%%%%%%%%%%%%%%%%%%%%%%%%%%%%%%

\subsection{Summary}

We have developed a rigorous pipeline to analyse the effects of FDM on the DES-Y1 data, and thus derived a new lower bound on the FDM particle mass combining DES-Y1 and \emph{Planck} data. Our pipeline uses likelihoods and samplers from \cosmosis. The linear power spectrum was computed with \acamb, while the non-linear power was modelled using the halo model as implemented in \HMC. \HMC\,is calibrated to simulations of CDM, including effects that lead to moderate suppression of the linear $P(k)$ on small scales, thus capturing some of the effects of FDM. We modify the halo model for FDM by introducing additional cut-offs in the HMF and halo concentration, seen in $N$-body simulations of FDM. Uncertainties on the non-linear model are propagated through our analysis pipeline by marginalising over nuisance parameters. We found that the additional uncertainty on the non-linear model of FDM did not affect our results, since the additional parameters of the model are uncorrelated with the cosmological parameters, in particular to the FDM mass, and are unconstrained by the data at the current level of precision. 

We also presented a bound on the FDM mass from the CMB alone, demanding that FDM is all of the DM, we find $\log_{10}(m/\text{eV})\geq -24.6 \quad \text{(95\% C.L.)}$ The combination of DES-Y1 and \emph{Planck} data leads to an improvement of this bound by approximately two orders of magnitude with respect to the CMB alone: $\log_{10} (m/\text{eV})\geq -23.0 \quad \text{(95\% C.L.)}$ The improvement is driven by the fact that DES data on small angular scales probes the cut-off in the linear power spectrum induced by FDM on scales inaccessible to the CMB. The CMB plays the role of fixing the primary cosmological parameters on large scales, leaving no room for degeneracy between the FDM induced cut-off and e.g. the power spectrum amplitude or spectral index in their effects on the small scales probed by DES. These results are shown in Fig.~\ref{fig:posterior_1d}.

A striking result of \HMC\,applied to FDM is that the one-to-two halo smoothing parameter leads to FDM \emph{enhancing} the power relative to CDM over a small range of scales in the quasi-linear regime and particle masses around $10^{-24}\text{ eV}$. We checked that this effect is largely within the calibrated range of the linear transfer function parameters of \HMC, within the redshift range probed by DES galaxies. When the FDM effect leaves the calibrated range, we tested that this does not drive constraints by comparing a variety of treatments for the smoothing parameter, and finding that our bound on the particle mass was unaffected. The increase in power can be explained qualitatively by noting that halo bias in models with reduced linear power increases on scales near the suppression scale, leading in some cases to slightly increased non-linear power on larger scales (``moving power around''). \HMC\,does not directly compute the halo bias, however effects such as this can be captured in the calibration to simulations by the phenomenological modifications to the basic halo model parameterised by an effective slope at the non-linear scale.  

In our main analysis, we chose to use only the shear correlation functions in the DES-Y1 data, $\xi_\pm$. This is because there are possible unaccounted for aspects of the galaxy bias in FDM in our model, which might call an analysis of the galaxy power spectrum, and galaxy-lensing cross correlation into question. However, we also presented an analysis including these data with a simple linear bias model for FDM, leading to a slight improvement in the lower bound on the particle mass to $\log_{10}(m/\text{eV})\geq -22.8 \quad \text{(95\% C.L.)}$ (see the Supplementary Material).

We also considered varying the statistical aspects of our analysis. We first compared different sampling techniques of MCMC and nested sampling, finding agreement between both. We next considered the effect of different priors on the FDM mass. Our main analysis adopts a log-flat prior between rough limits imposed by previous analyses, and set by the expected sensitivity of the data. We compared these results to the case with a uniform prior on the inverse mass, which allows finite prior volume at the CDM limit. Accounting for the different prior volumes, the results are consistent (as expected).

Having developed such a rigorous analysis, we can be confident that the presented bound on the FDM mass is statistically and theoretically robust.

\subsection{Comparison to Other Results}

Our model approximates FDM as having vanishing self-interactions, i.e. scalar potential $V(\phi)\approx m^2\phi^2/2$. Axions are expected to have periodic potentials, and so if the FDM is an ultra-light axion, its self interactions are described by the potential, $V=m^2f_a^2[1-\cos{(\phi/f_a)}]$, where $f_a$ is the axion decay constant. \cite{LinaresCedeno:2020dte} presented constraints on ULAs including such interactions parameterised by a coupling constant $\lambda=3M_{pl}^2/f_a^2$, and showed that the limit on $m$ is not affected significantly by the self-interaction strength for parameter values where the implementation of linear perturbation theory is numerically reliable. 

Using the CMB,  \cite{LinaresCedeno:2020dte} find the bound $\log_{10}(m/\text{eV})>-23.99$ at 95\% C.L., which is slightly stronger than our bound form the CMB alone. However, \cite{LinaresCedeno:2020dte} took a maximum value of $\log_{10} (m/\text{eV})=-16$ in their log-flat prior. This is far beyond the sensitivity of the CMB, and so their limit is dominated by the choice of prior. 

Many complementary cosmological probes can also be used to measure the FDM mass and density fraction relative to CDM including galaxy formation~\citep{Schive:2015kza,Bozek:2014uqa,Corasaniti:2016epp}, X-ray observations~\citep{Maleki2020:InvestigationOf}, the Lyman-$\alpha$ effective opacity~\citep{Sarkar2021:UsingRedshift}, and the galaxy power spectrum multipoles (which can be used in conjunction with CMB data)~\citep{Lague:2021frh}. Upcoming cosmological surveys will further improve sensitivity to FDM. Forecasts from line intensity mapping surveys project that the ULA fraction could be measured at the percent level~\citep{Bauer2021IntensityMapping}, and is sensitive to masses on the order of $10^{-22}\text{ eV}$ and below. There is similar sensitivity for future CMB experiments, especially so-called ``high definition CMB''~\citep{Hlozek:2016lzm,Sehgal:2019ewc}. Measurements of the cluster pairwise velocity dispersion through the kinetic Sunyaev-Zel'dovich (kSZ) effect could probe ULA mass fractions of $\sim 5\%$ in a window near $m\sim 10^{-27}~{\rm eV}$, while the high-$\ell$ Ostriker-Vishniac anisotropies could reach mass fractions of $0.1\%$ near $m\sim 10^{-27}~{\rm eV}$, and is sensitive to pure FDM up to masses as high as $m\sim 10^{-22}~{\rm eV}$~\citep{Farren:2021jcd}. Future pulsar timing array measurements could probe axion masses around $10^{-22}$ eV~\citep{Porayko2018PTA}, and percent level fractions for $m\approx 10^{-23}\text{ eV}$~\citep{Khmelnitsky:2013lxt}. 

Additional bounds on the FDM mass can be obtained using astrophysical data on the local distribution of DM. Rotation curves and velocity dispersions of many types of galaxies are affected by the ``solitonic core'' formed in FDM density profiles  ~\citep{schive2014cosmic,Marsh:2015wka,Chen2017JeansAnalysisDwarf,Gonzalez-Morales:2016yaf,Wasserman2019SpatiallyResolved, Hayashi2021NarrowingMassRange,Bar:2018acw}. We note, however, that these probes are affected by important systematic errors. Indeed it has been shown that stars, gas and black holes have significant impacts on the small scale ULA density profile~\citep{Veltmaat2020BaryonDrivenGrowth,Davies2020FDMSolitonCores,2018MNRAS.478.2686C}. Stringent constraints have been obtained using black hole superradiance~\citep{Arvanitaki:2009fg,Arvanitaki2011ExploringStringAxiverse,Stott:2018opm,StottUltralightBosonic} which excludes narrow bands in $m$ above the region where FDM has significant effects on cosmology and galaxy formation \citep[the black hole in M87 reaches furthest down into the FDM range,][]{Davoudiasl:2019nlo} . Heating of stars by small scale fluctuations of FDM also provides a probe of the particle mass~\citep{Hui:2016ltb}. Heating of the Milky Way disk leads to the limit $m\gtrsim 10^{-22}\text{ eV}$~\citep{Church2019HeatingMilkyWay}, while the survival of the Eridanus-II star cluster excludes a range of masses near $10^{-20}\text{ eV}$~\citep{Marsh:2018zyw,Chiang:2021uvt}. The Milky Way satellite population, including Eridanus-II~\citep{Marsh:2018zyw}, and the satellites surveyed by DES~\citep{Nadler2019ConstraintsDMMicrophysics}, appears to disfavour $m\lesssim 10^{-21}\text{ eV}$, although it has been argued that this bound can be avoided if the true masses are underestimated, and ultra faint galaxies are in fact tidally stripped and isolated solitons~\citep{Chiang:2021uvt,Broadhurst:2019fsl}. 

Finally, we note the recent results of \cite{Blum2021:GrravitationalLensing} and \cite{Allali2021:DarkSector}, who find that the introduction of a partial ULA dark matter component may help alleviate the various ``tensions'' in recent cosmological data combinations. \cite{Blum2021:GrravitationalLensing} finds that $10^{-25}$~eV ULAs composing 10\% of the dark matter could alleviate the $H_0$ tension from gravitational lensing. \cite{Allali2021:DarkSector} proposes a mixed DM model with ULAs paired with a decaying dark energy component that can simultaneously resolve the $H_0$ and $\sigma_8$ tensions.

\subsection{Looking Forward}

On small angular scales, $\theta\lesssim \mathcal{O}(1)$~arcmin, baryonic effects in the form of Active Galactic Nucleus (AGN) feedback can affect weak lensing correlation functions. \HMC\,has been calibrated to simulations of AGN feedback for a wide range of cosmologies, and in the weak lensing analysis analysis of KiDS in \cite{Joudaki:2019pmv} the uncertainty in the AGN feedback model is marginalised over. AGN feedback is not expected to be precisely the same for CDM and FDM. The DES-Y1 analysis masks small angular scales in order to avoid uncertainty due to AGN feedback (indicated by the grey regions in Figs.~\ref{fig:shear_powerspectra}, \ref{fig:shear_plusmminus_models_data} and \ref{fig:extra_models_data}), and so our present analysis is not affected by the model of AGN feedback. Taking full advantage of weak lensing on small angular scales to constrain the nature of DM requires simulations including AGN feedback beyond CDM in order to calibrate e.g. \HMC\,and emulators.

If such calibrations were performed, how could weak lensing tests of FDM improve in the future? We can estimate this using the bounds to the effective linear power spectrum~\citep{Tegmark:2002cy,Chabanier:2019eai}. \cite{Tegmark:2002cy} showed that the effective linear theory wavenumber $k\propto 1/\theta$ for weak lensing correlation functions measured on an angular scale $\theta$. In our analysis, the smallest value of $\theta$ was $\theta\approx 4$ arcmin in $\xi_+$.  The smallest scales used in the DES-Y3 analysis~\citep{DES:2021wwk} are around 1 arcmin. Using again that $k_{\rm J,eq}\propto m^{1/2}$ (Eq.~\ref{eqn:kj_eq}), substantial improvement in ULA constraints is possible with existing data, to $m\sim 10^{-22}\text{ eV}$ (at the time of writing the DES-Y3 likelihood is not yet available). The \emph{Euclid} angular resolution is around 0.1 arcmin~\citep{2011arXiv1110.3193L}, which by the same logic might be sensitive to $m\sim 10^{-20}\text{ eV}$, comparable to the Lyman-alpha forest constraints of \cite{Rogers:2020ltq}.

When smaller scales are included in weak lensing analyses of FDM, the effects on the power spectrum from the models of $c(M)$ and $n(M)$ will become increasingly important. A rigorous analysis of such data will require  the tools we have developed to account for systematic uncertainty in the low mass behaviour of $c(M)$ and $n(M)$. We also advocate more in depth studies of the one-to-two halo transition region in simulations of FDM and mixed models of CDM and FDM.

%
%%%%%%%%%%%%%%%%%%%%%%%%%%%%%%%%%%%%%%%%%%%%%%%%%%%%%%%%%%%%%%%%%%%%%%%%%%%%%%%%%
%%%%%%%%%%%%%%%%%%%%%%%%%%%%%%%%%%%%%%%%%%%%%%%%%%%%%%%%%%%%%%%%%%%%%%%%%%%%%%%%%
\section*{Acknowledgments}
%%%%%%%%%%%%%%%%%%%%%%%%%%%%%%%%%%%%%%%%%%%%%%%%%%%%%%%%%%%%%%%%%%%%%%%%%%%%%%%%%
%%%%%%%%%%%%%%%%%%%%%%%%%%%%%%%%%%%%%%%%%%%%%%%%%%%%%%%%%%%%%%%%%%%%%%%%%%%%%%%%%
MD and DJEM were supported at the University of G\"{o}ttingen by the Alexander von Humboldt Foundationa and the German Federal Ministry of Education and Research. DJEM is supported at King's College London by an Ernest Rutherford Fellowship of the Science and Technologies Facilities Council (UK). RH is a CIFAR Azrieli Global Scholar in the Gravity and the Extreme Universe Program, and acknowledges funding from the NSERC Discovery Grants program, the Alfred P. Sloan Foundation and the Connaught Fund. The Dunlap Institute is funded through an endowment established by the David Dunlap family and the University of Toronto. The authors at the University of Toronto acknowledge that the land on which the University of Toronto is built is the traditional territory of the Haudenosaunee, and most recently, the territory of the Mississaugas of the New Credit First Nation. They are grateful to have the opportunity to work in the community, on this territory. DG acknowledges support in part by NASA ATP grant 17-ATP17-0162 and the Provost's office at Haverford College. We would like to thank the following colleagues for useful discussions: Sebastian Hoof, Shahab Joudaki, Alex van Engelen. We thank Aaron Ludlow for supplying code to reproduce the work of \cite{Ludlow:2016ifl}. MD thanks members of the Munich cosmology group for helpful suggestions. We analysed some of our results using \textsc{ChainConsumer}~\citep{Hinton2016}, and made use of the following open source libraries: \textsc{numpy}~\cite{numpy}, \textsc{matplotlib}~\cite{matplotlib}. Our \textsc{cosmosis} analysis relies, in addition to what is cited in the text, on the following works: \cite{2012MNRAS.424.1647K,Bridle:2007ft,Kilbinger:2008gk}. The \emph{Planck} data used in this article was accessed through the likelihoods in \textsc{cosmosis}, the original data are available here: \href{https://wiki.cosmos.esa.int/planck-legacy-archive/index.php/CMB_spectrum_%26_Likelihood_Code}{Planck Legacy Archive}.

This project used public archival data from the Dark Energy Survey (DES), based on \cite{2018ApJS..239...18A,2018PASP..130g4501M,2015AJ....150..150F}, \href{https://des.ncsa.illinois.edu/releases/y1a1}{Data Products}. Funding for the DES Projects has been provided by the U.S. Department of Energy, the U.S. National Science Foundation, the Ministry of Science and Education of Spain, the Science and Technology FacilitiesCouncil of the United Kingdom, the Higher Education Funding Council for England, the National Center for Supercomputing Applications at the University of Illinois at Urbana-Champaign, the Kavli Institute of Cosmological Physics at the University of Chicago, the Center for Cosmology and Astro-Particle Physics at the Ohio State University, the Mitchell Institute for Fundamental Physics and Astronomy at Texas A\&M University, Financiadora de Estudos e Projetos, Funda{\c c}{\~a}o Carlos Chagas Filho de Amparo {\`a} Pesquisa do Estado do Rio de Janeiro, Conselho Nacional de Desenvolvimento Cient{\'i}fico e Tecnol{\'o}gico and the Minist{\'e}rio da Ci{\^e}ncia, Tecnologia e Inova{\c c}{\~a}o, the Deutsche Forschungsgemeinschaft, and the Collaborating Institutions in the Dark Energy Survey.
The Collaborating Institutions are Argonne National Laboratory, the University of California at Santa Cruz, the University of Cambridge, Centro de Investigaciones Energ{\'e}ticas, Medioambientales y Tecnol{\'o}gicas-Madrid, the University of Chicago, University College London, the DES-Brazil Consortium, the University of Edinburgh, the Eidgen{\"o}ssische Technische Hochschule (ETH) Z{\"u}rich,  Fermi National Accelerator Laboratory, the University of Illinois at Urbana-Champaign, the Institut de Ci{\`e}ncies de l'Espai (IEEC/CSIC), the Institut de F{\'i}sica d'Altes Energies, Lawrence Berkeley National Laboratory, the Ludwig-Maximilians Universit{\"a}t M{\"u}nchen and the associated Excellence Cluster Universe, the University of Michigan, the National Optical Astronomy Observatory, the University of Nottingham, The Ohio State University, the OzDES Membership Consortium, the University of Pennsylvania, the University of Portsmouth, SLAC National Accelerator Laboratory, Stanford University, the University of Sussex, and Texas A\&M University.
Based in part on observations at Cerro Tololo Inter-American Observatory, National Optical Astronomy Observatory, which is operated by the Association of Universities for Research in Astronomy (AURA) under a cooperative agreement with the National Science Foundation.

\section*{Data Availability Statement}

The data underlying this article, in the form of chains, will be shared on reasonable request to the corresponding author.
%
%%%%%%%%%%%%%%%%%%%%%%%%%%%%%%%%%%%%%%%%%%%%%%%%%%%%%%%%%%%%%%%%%%%%%%%%%%%%%%%%%
%%%%%%%%%%%%%%%%%%%%%%%%%%%%%%%%%%%%%%%%%%%%%%%%%%%%%%%%%%%%%%%%%%%%%%%%%%%%%%%%%
%%%%%%%%%%%%%%%%%%%%%%%%%%%%%%%%%%%%%%%%%%%%%%%%%%%%%%%%%%%%%%%%%%%%%%%%%%%%%%%%%

\bibliographystyle{mnras}
\bibliography{ref}
\appendix
%%%%%%%%%%%%%%%%%%%%%
\section{DES-Y1 3x2pt}\label{appendix:extra_data}
\begin{figure*}
\begin{center}
 \includegraphics[width=1\textwidth]{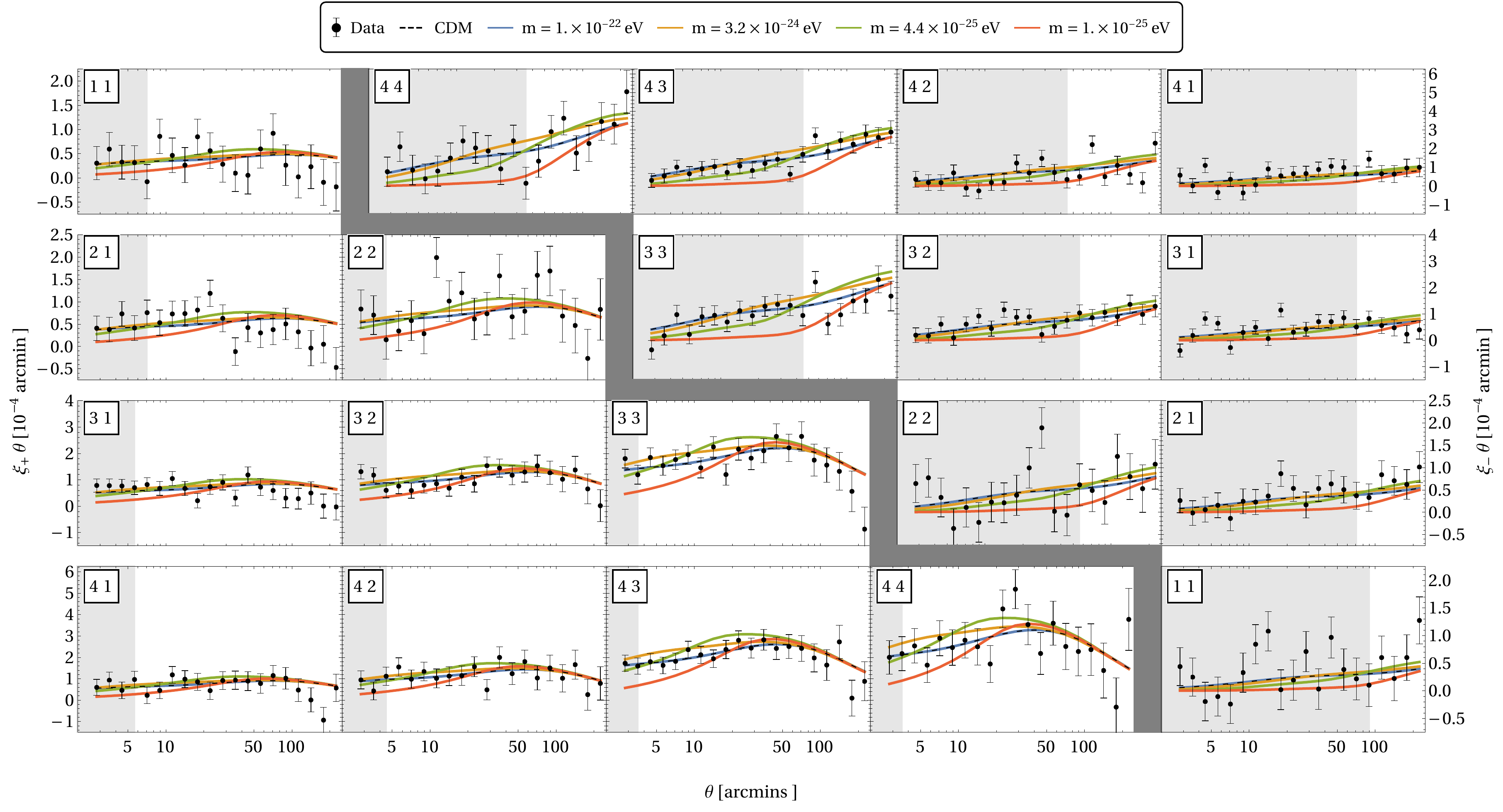}\\
 \caption{Cosmic shear correlation function between all combinations $i\,j$ of source bins, as indicated in the top left corner of each panel.  We display the $\xi_+$ observable on the left hand side and the $\xi_-$ observable on the right hand side (note the different order of bin combinations shown on each side). We show DES-Y1 data points in black, marking the square-roots of the diagonal entries of the covariance matrix as vertical bars. We compare the CDM predictions (black, dashed lines) to the ULA predictions taking four different values for the axion mass parameter $m_{Ax}$ within our prior range (solid, colour-coded). The grey shaded regions mask angular scales excluded from the analysis due to modelling uncertainties. The case $m=10^{-22}\text{ eV}$ is indistinguishable from CDM over the scales shown. }
 \label{fig:shear_plusmminus_models_data}
\end{center}
\end{figure*}
\begin{figure*}
\begin{center}
\includegraphics[width=\textwidth]{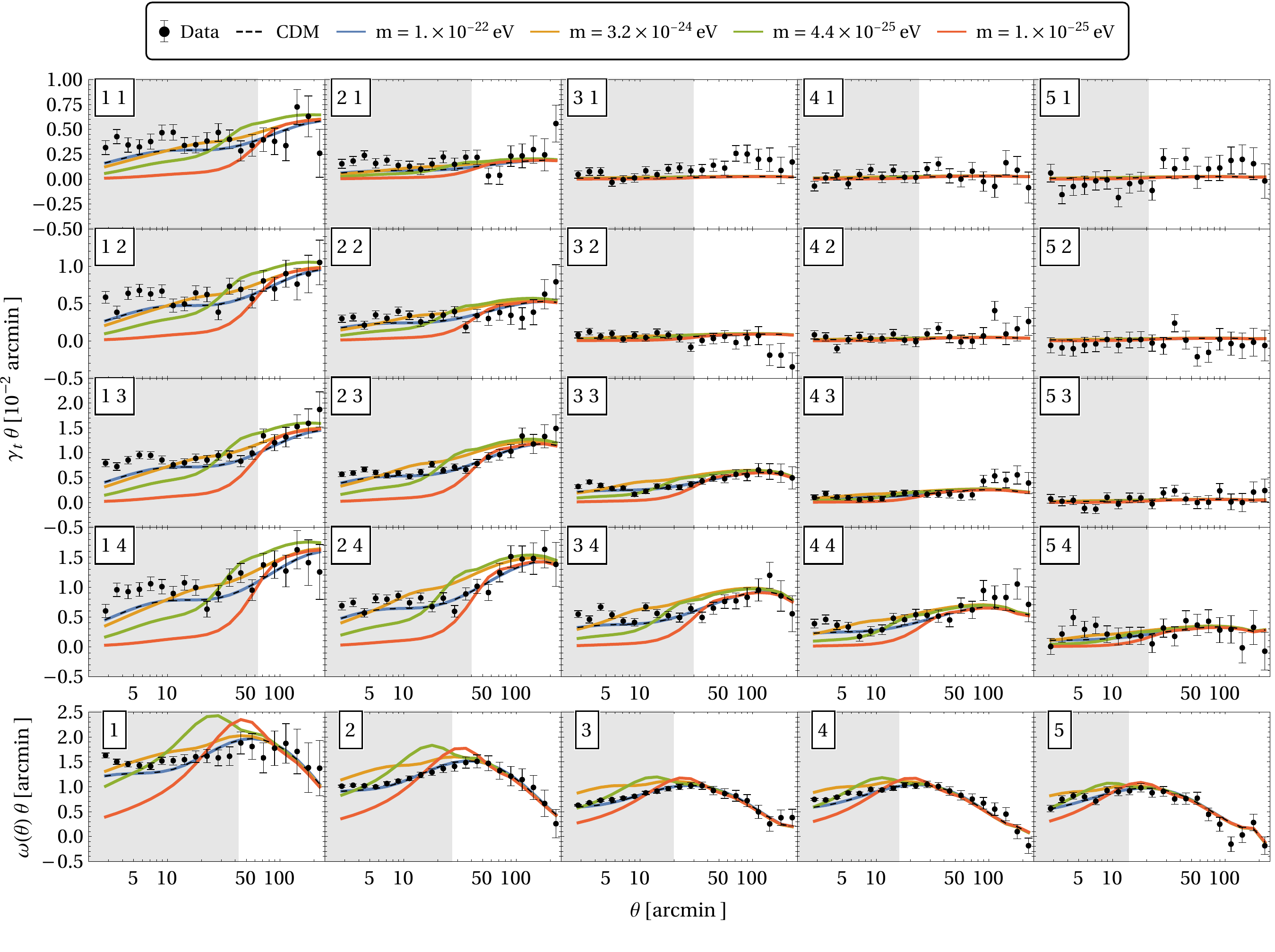}
 \caption{The $\gamma$ and $w$ correlation functions that form the ``3x2pt'' observable when combined with the $\xi_{\pm}.$ The results obtained when using the full data set are compared to our baseline analysis in Fig.~\ref{fig:shear_vs_all}, and show a modest improvement in the bound on the axion mass.\label{fig:extra_models_data}}
\end{center}
\end{figure*}

\begin{figure*}
    \centering
    \includegraphics[width=.7\textwidth]{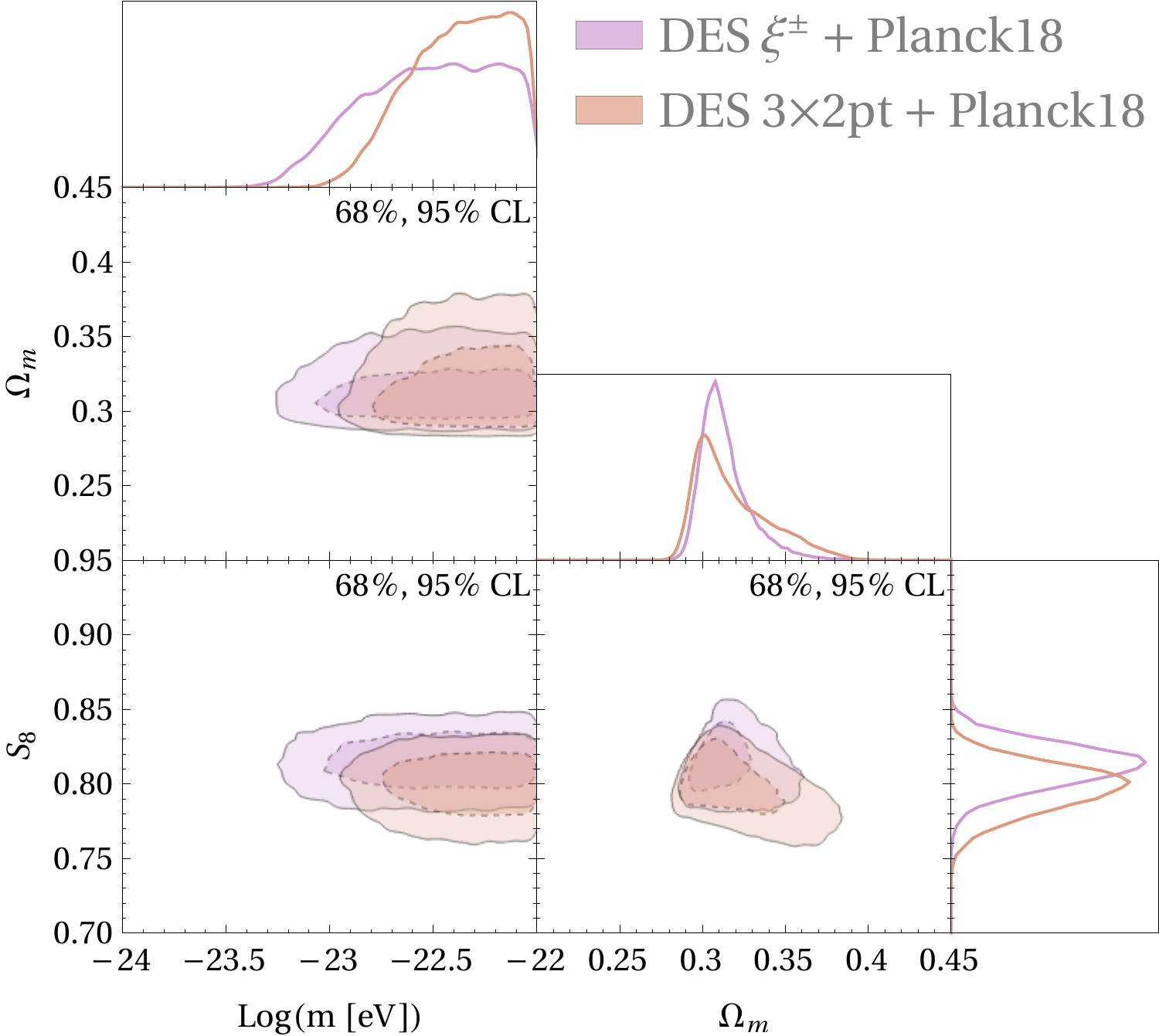}
    \caption{Posterior density for the DM particle mass $m$, the mass fluctuation parameter $S_8$, and the matter density $\Omega_m$ for different combinations of data sets. DES shear-only data set (pink) and the complete 3x2pt data set (red). Note that we changed the range of the axis compared to Fig.~\ref{fig:axion_cosmo} for better visibility.}
    \label{fig:shear_vs_all}
\end{figure*}

The DES-Y1 data also contains the galaxy power spectrum in the form of the correlation function, $w$, and the cross correlation between shear and galaxy number, $\gamma_t$, leading to the complete ``3x2pt'' dataset. These data (shown in Figure~\ref{fig:extra_models_data}, along with the predictions for CDM and FDM given reference cosmological and bias parameters) are not included in our baseline analysis, since $w$ and $\gamma_t$ depend on the model of galaxy bias, for which a linear, scale independent model may not suffice for FDM \citep[as discussed briefly in e.g.][]{Hlozek:2014lca}. Nonetheless, we performed our analysis assuming standard linear bias for the full DES-Y1 3x2pt plus \emph{Planck} data. Following \cite{Abbott:2017wau} we note that the galaxy power $P_{\rm gal}(k)$, and thus $w$, is quadratic in the galaxy bias, while $\gamma_t$ has a linear response. Thus in the 3x2pt analysis it is possible to measure the binned linear bias parameters $b_i$. The results of this analysis are shown in Fig.~\ref{fig:shear_vs_all} for the posterior on the FDM mass, and correlation with the power spectrum amplitude and matter density parameter.

%%%%%%%%%%%%%%%%%%%%%%%%%%%%%%%%%%%%%%%%%%%%%%%%%%%%%%%%%%%%%%%%%%%%%%%%%%%%%%%%%
%%%%%%%%%%%%%%%%%%%%%%%%%%%%%%%%%%%%%%%%%%%%%%%%%%%%%%%%%%%%%%%%%%%%%%%%%%%%%%%%%
\section{Beyond CDM Effects Not Included in Our Halo Model}\label{appendix:halo_omissions}

%%%%%%%%%%%%%%%%%%%%%%%%%%%%%%%%%%%%%%%%%%%%%%%%%%%%%%%%%%%%%%%%%%%%%%%%%%%%%%%%%
\subsection{Solitons in the Power Spectrum}
\label{appendix:solitons}
%%%%%%%%%%%%%%%%%%%%%%%
\begin{figure*}
 \begin{center}
 \includegraphics[width=.9\textwidth]{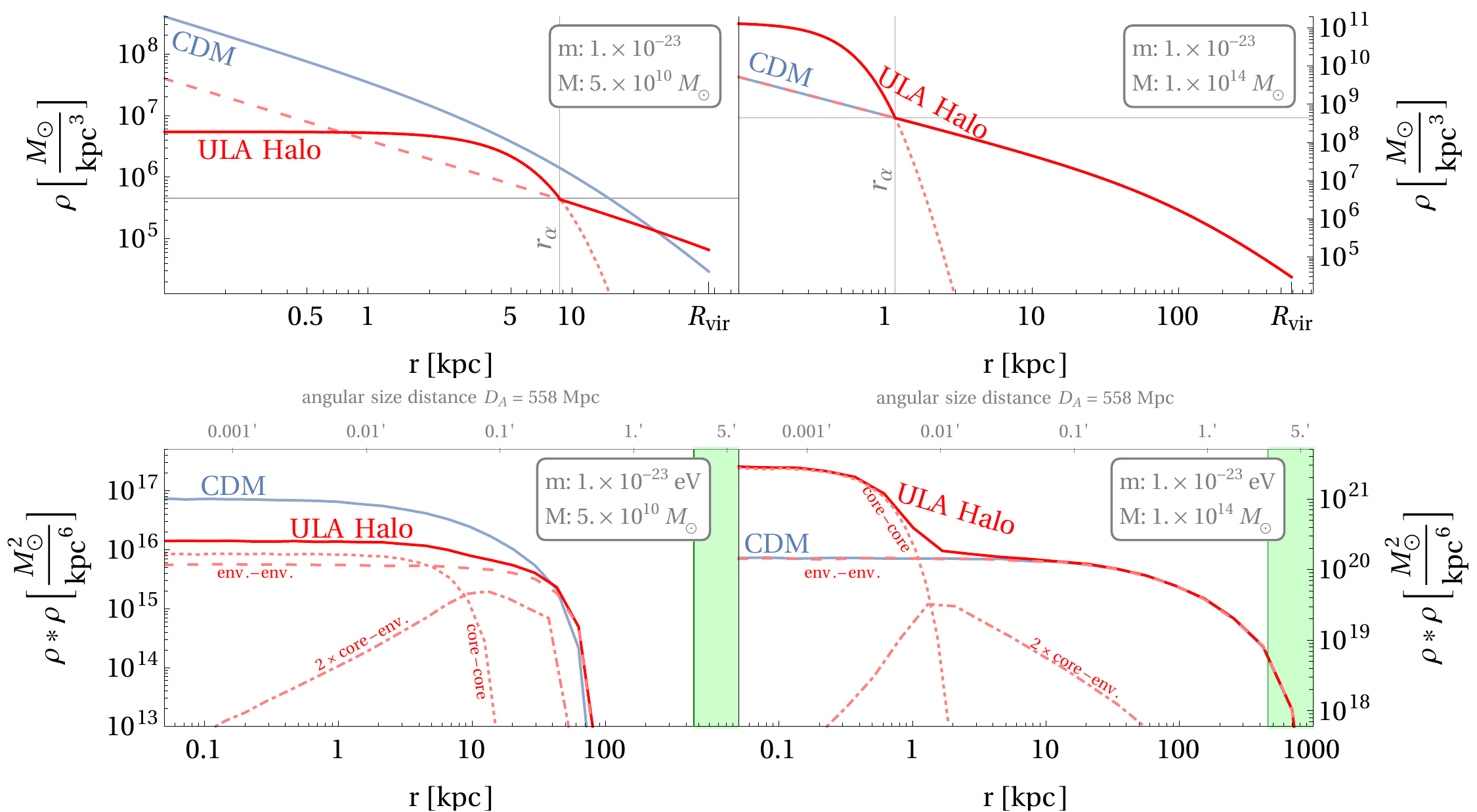}
\end{center}
\caption{Halo density profiles  $\rho$ (top) and auto-correlation functions $\rho\star\rho$ (bottom) for two different halo masses, for FDM mass $m=10^{-23}\text{ eV}$ (red) and CDM (blue). We compare a less massive halo $M=5\times10^{10}\,\mathrm{M_\odot}$ (left) to a more massive halo $M=1\times10^{14}\,\mathrm{M_\odot}$ (right). In the case of CDM, we show the NFW profile given in Eq.~\eqref{eq:NFW}, which corresponds to setting the parameter $\eta=0$ in \HMC\,(i.e. no halo bloating). In the case of FDM we show a piece-wise profile with a solitonic core and an envelope parameterised by a modified NFW profile, as given in Eq.~\eqref{eq:ULAprofile}. The matching radius $r_\alpha$ is set to $r_\alpha=2$ in the left-hand column, similar to the values found in~\protect\cite{10.1093/mnras/sty3190}. In the right-hand column, we chose $r_\alpha=1.2$ which leads to the same concentration-mass relation for the NFW profile and the profile including the solitonic core. In the top row, we show in dotted (dashed) what the ULA profile would look like if we extrapolated the core (envelope) profile beyond the matching radius. In the bottom row, we depict the full correlation function in solid. For FDM we show in dotted, dashed, and dot-dashed the contributions of the core-core, envelope-envelope and core-envelope correlation functions to the full result (see text for details).
To guide the eye, we display the angular scale of the halo density auto-correlation for angular diameter distance $D_A=558\text{ Mpc}$ (roughly 5\% of the galaxies in the DES-Y1 sample, at $z=0.15$). The green shaded region shows the angular resolution of DES.}
\label{fig:EffectOfCore}
\end{figure*}
%%%%%%%%%%%%%%%%%%%%%%%%%%%%%%%
As described in section \ref{subsec:axionHM}, the particular physical properties of FDM impact the distribution of matter on different scales. On very small scales, matter consisting of FDM forms a stable compact object called a ``soliton" or ``axion star" \citep[e.g.][]{Ruffini:1969qy,Seidel:1991zh,Chen:2020cef,Helfer:2016ljl,schive2014cosmic,Levkov:2018kau}. Solitons appear as an additional, core-like central structure in the density profile of FDM halos. In our implementation of the halo model, we chose not to incorporate solitons. In this section we justify this approach and assess the possibility of discovering solitons in future lensing surveys. 

Solitons are found in numerical simulations of FDM haloes solving the fully non-linear Schr\"{o}dinger-Poisson equations, as first shown by \cite{schive2014cosmic}. Soliton density profiles depend on $m$ and the soliton mass, $M_{\rm sol}$ (equivalently, the core radius), and can be fit very well by the the ground state solution of the  Schr\"odinger-Poisson equation. The solitonic core is found to be embedded in an envelope halo which can be parameterised by a CDM-like density profile. We follow the phenomenological approach of \cite{Marsh:2015wka,10.1093/mnras/sty3190} and parameterise the density profile of a FDM halo by the piece-wise equation
\begin{align}
    \rho_{\rm FDM}(r) = \begin{cases} 
           \rho_{\rm core}(r)& 0\leq r\leq r_\alpha \\
          \rho_{\rm env} & r_\alpha< r\leq R_{vir},
       \end{cases}\label{eq:ULAprofile}
\end{align}
with~\citep{schive2014cosmic}
\begin{align}
    \rho_{\rm core}(r)=\frac{\rho_c}{\left[1+0.091\left(\frac{r}{r_c}\right)^2\right]^8}
\end{align}
and the central density $\rho_c$ given by 
\begin{align}
 \rho_c=2.94\times10^6\mathrm{M_\odot kpc}^{-3}(M/10^9\mathrm{M_\odot})^{4/3}(m/10^{-22}\text{ eV})^2\, .
\end{align}
The central density defined above is specified in terms of the halo mass assuming the ``core-halo mass relation''~\citep{Schive:2014hza}, which fixes the soliton mass, and thus also the core radius, $r_c$, in terms of the total halo mass $M$. The relevant equations can be found in \cite{Schive:2014hza}.

The radius corresponding to $\rho_{\rm core}(r)=\rho_c/2$ is defined as the core radius $r_c$. The transition radius $r_\alpha$ is modelled as $r_\alpha=\alpha\,r_c$, where the parameter $\alpha$ is a number that increases from $\sim2$ for small haloes to $\sim 4$ for larger halos \citep{10.1093/mnras/sty3190}, again assuming the core-halo mass relation. The envelope profile $\rho_{\rm env}$ can be described by the same functional form as a CDM halo. However, to ensure continuity at the matching radius $r_\alpha$ the respective parameters of the envelope profile, i. e. the envelope mass $M_{\rm env}(r_\alpha)$ and the concentration parameter $c_{\rm env}(M,r_\alpha)$ need to be adapted accordingly. 
Note that for larger halos, we can chose $r_\alpha$ such that the envelope  concentration $c_{\rm env}(M,r_\alpha)$ is equal to the concentration $c_{\rm NFW}$ of a pure NFW halo of the same total mass $M$. If, however, for a lighter halo $\rho_{\rm core}(r,M)<\rho_{\rm NFW}(r,M)$ for all radii $r$, we can in general find no $r_\alpha$ such that the concentration-mass relation is conserved for the halo including the solitonic core. This implies that in addition to the peculiar feature of a core, in addition, we expect that smaller halos have a different outer profile, making them even more distinct from a NFW halo of the same mass.

We show the resulting profiles in Fig.~\ref{fig:EffectOfCore} for two different halo masses $M=5\times10^{10}\,\mathrm{M_\odot}$ and $M=1\times10^{14}\,\mathrm{M_\odot}$ in comparison to the respective profiles of CDM halos of the same mass. For both cases, the solitonic core is a distinct feature setting the FDM halo profile apart from the CDM halo profile. Therefore, solitons are potentially very valuable signatures in the quest for the discovery of FDM. Solitons are, however, very compact objects with a transition radius around $r_\alpha\sim\mathcal{O}(\text{1-10\,kpc})$ which holds true for smaller and larger halos alike. This is due to the scaling relations $ r_c/R_{\rm vir}\propto M^{-2/3}$ and $M_{\rm sol}/M\propto M^{-2/3}$ implying that the relative core-size decreases for more massive halos. Consequently, the contribution of the core to the overall profile and total mass becomes progressively less significant for more massive halos \citep{10.1093/mnras/sty3190,schive2014cosmic}. 

This implies that for the relatively massive halo in the right-hand column of Fig. \ref{fig:EffectOfCore}, the envelope profile is very similar to the respective profile of a CDM halo of the same mass, when picking the scaling radius in an appropriate range~\citep{10.1093/mnras/sty3190}. We chose a scaling radius of $r_\alpha=1.2$ such that concentration-mass relation for the large halo in the right column of Fig. \ref{fig:EffectOfCore} is the same assuming a solitonic core and assuming a pure NFW profile, motivated by results reported from simulations~\cite{schwabe2021deep}. In this case, the differences between the FDM and CDM halo profiles are virtually invisible except for the small core region with $r<r_\alpha~1\,\mathrm{kpc}$.  For the less massive halo in the left-hand column of Fig. \ref{fig:EffectOfCore}, the core takes up a larger relative fraction of the total halo size and mass. Therefore, in this case the envelope profile deviates significantly from the respective profile of a CDM halo of the same mass and virial radius, and we need to adapt the concentration-mass relation, as explained above. In this case, for illustration, we chose a rather arbitrary scaling radius $r_\alpha=2$.

We now turn to the possible effect of a soliton core on the DES observables. The profiles add to the non-linear power spectrum model in the form of the halo density autocorrelation $\rho_h\star\rho_h$, weighted by the HMF, in the 1h-term given in Eq.~\eqref{eq:1h}. Using Eq.~\eqref{eq:ULAprofile}, we can factorise the autocorrelation as $\rho_h\star\rho_h=\rho_{core}\star\rho_{core}+2\rho_{core}\star\rho_{env}+\rho_{env}\star\rho_{env}$. In the bottom row of Fig.~\ref{fig:EffectOfCore}, we show the halo density autocorrelation. We observe that the core-core correlation dominates the total correlation function up to $r\lesssim r_\alpha$. The envelope-envelope correlation dominates for $r\gtrsim r_\alpha$, while the core-envelope term is subdominant. We notice that the FDM halo density autocorrelation for $M=5\times10^{10}\,\mathrm{M_\odot}$ differs significantly from the CDM counterpart on its support $r\leq 2R_{vir}$, while the autocorrelations agree up to a few percent for the heavy halo with $M=1\times10^{14}\,\mathrm{M_\odot}$, except for very small $r\lesssim r_\alpha$. 

However, these differences in the halo auto-correlation between FDM and CDM for $m=10^{-23}\text{ eV}$ cannot be resolved within the DES survey. This is because the minimum angular bin is $\theta=2.8'$, which we can translate to the physical distance $r$ in the autocorellation by means of the  angular diameter distance $D_A=r/\theta$. For the source distribution in our sample, we in general expect the resolution to be highest for the nearest galaxies. We show in Fig.~\ref{fig:EffectOfCore} the apparent angular size at redshift $z=0.15$, with $D_A=558\text{ Mpc}$ for our best fit cosmology (less than 5\% of galaxies in the lowest bin of the DES data have a lower redshift).  The corresponding region in the autocorrelation is marked as a green band. We observe in Fig.~\ref{fig:EffectOfCore} that only the density correlation in the outer regions of large halos can be resolved, where the density profile is dominated by the envelope, which is very similar to the corresponding CDM case. This justifies our approach  to use the unmodified CDM halo profile also for FDM halos in our implementation of the halo model, and in our analysis: the effects of solitons appear on small angular scales that are not resolved by DES-Y1. 

So far, we have argued that the minimum angular resolution of the DES survey does not allow to measure the halo density autocorrelation  of small halos for which the profile deviates significantly from the CDM counterpart. We conclude with the remark that even assuming much higher angular resolution, a detection of FDM features in halos using weak lensing is very challenging, because small FDM halos are suppressed by the additional cut-off in the HMF, as discussed in Section~\ref{sec:nl_power}. We have considered the effects of solitons for $m=10^{-23}\text{ eV}$, at the upper edge of our  posterior distribution. For larger masses, solitons become smaller, and thus even harder to observe, while for lower masses solitons would lead to stronger deviations from CDM predictions in the observables. 

\subsection{Correlation of the Smooth Component}

The halo model, as described in Section~\ref{subsec:HM}, implicitly assumes that all DM is bound in halos. This assumption is valid for models such as CDM with no free streaming scale, where the variance of fluctuations diverges as the length scale in the filter $R\rightarrow 0$. For models with a free streaming scale, including mixed DM with CDM and massive neutrinos, with pure WDM, or pure FDM (or any admixture of these models), there is cut-off scale below which some of the DM does not reside in halos. As described in Section~\ref{subsec:axionHM}, we implement this effect for FDM by modifying the HMF such that it is significantly suppressed below the cut-off scale. With our choice of the HMF Eq.~\eqref{eq:corr_HMF_ULA}, the fraction of the matter density bound in halos, $f_{\rm halo}$, can be calculated according to
\begin{align}
    f_{\rm halo} = \frac{1}{\bar{\rho}}\int dM\, M n(M)=\frac{\bar\rho_{\rm h}}{\bar{\rho}}\lesssim1.
    \label{eq:f_halo}
\end{align}
By accounting for the cut-off in the HMF we consistently model the one-halo term in our halo model. 
However, in addition, there now exists a component of DM unbound to halos, known as the smooth component. This implies that in addition to the two-halo term, we need to consider the  auto-correlation of the smooth component as well as the cross-correlation of the smooth component and the matter contained in halos.

The halo model can be adapted to account for such a smooth component: e.g. for CDM and neutrinos see \cite{Massara:2014kba}, while for pure WDM see \cite{Smith:2011ev}. The power spectrum with a smooth component is:
\be
P(k) = (1-f_{\rm halo})^2P_{\rm ss}+2(1-f_{\rm halo})P_{\rm sh}+P_{\rm 2h}+P_{\rm 1h}\, ,
\ee
where $s$ denotes the smooth component, while $h$ denotes the halo component, and we calculate the one-halo term $P_{\rm 1h}$ and the two-halo $P_{\rm 2h}$ term with respect to $n(M)$ \emph{including the small scale cut-off}. 
Note that we choose to normalise power spectra $P_{\rm 2h}$ and $P_{\rm 1h}$ as well as the cross-power spectrum $P_{\rm sh}$ with respect to the total mean matter density $\bar{\rho}$, not the matter density in halos $\rho_{\rm h}$, which eliminates the pre-factors of $f_{\rm halo}$ as compared to e.g.~\cite{Smith:2011ev}.  As in Section~\ref{subsec:HM}, we assume that halos are biased tracers of the linear matter density. It is furthermore reasonable to assume that also the smooth component traces the linear matter density~\citep{Smith:2011ev}, and that both the smooth and the halo component can be related to the linear field through the linear bias terms $b_{\rm s}(M)$ and $b_{\rm h}(M)$, respectively.

In the limit that the bias is ignored, the smooth-smooth, smooth-halo, and two-halo terms in the power spectrum all add up to equal the linear power spectrum regardless of the value of $f_{\rm halo}$~\citep{Smith:2011ev}. Thus, in the limit that the bias is neglected in the two halo term, one can neglect the smooth component correlations. \HMC\,neglects the bias since it is not necessary to compute the non-linear within the desired accuracy: the one-halo term dominates before the bias has a significant effect. Thus it is consistent within the approximations used by \HMC\,to neglect the smooth component correlations of FDM, as long as the one-halo term has the correct cut-off. Furthermore, we showed that within the accuracy of DES-Y1 the HMF cut-off does not affect our lower limit to $m$ (Fig.~\ref{fig:marg_HM_params}), which further reinforces the claim that smooth component correlations can be neglected within the desired accuracy.

\subsection{Relativistic Corrections}

ULAs are treated in our work at early times fully relativistically in linear perturbation theory within \acamb. The energy density and effective pressure contains contributions from the kinetic, gradient, and potential energy density of the FDM field, which all enter into the computation of the linear power spectrum. 

Our modifications to the halo model, i.e. the cut-offs in $n(M)$ and $c(M)$ are fit to $N$-body simulations, which take the linear power spectrum as input, but treat the DM entirely non-relativistically. The next level of improved approximation to FDM physics on non-linear scales involves the introduction of the quantum pressure term, and solution of the Schr\"{o}dinger-Poisson equations. The equations differ from an $N$-body model, since the equation of motion for the field contains gradient energy, an effective smoothing scale in the Vlasov equation~\citep{1993ApJ...416L..71W,Uhlemann:2014npa}. The quantum pressure term is expected to lead to extra suppression in $n(M)$ and $c(M)$ in addition to what is observed in $N$-body simulations (although no large enough simulations have been performed including this physics to make a definitive measurement), and we have accounted for this in our model for systematic uncertainty in $n(M)$ and $c(M)$. Quantum pressure also leads to the formation of solitons on small scales in FDM halos, which as we have argued above can be neglected on scales relevant to the DES-Y1 shear correlation.

There are, however, aspects of quantum pressure which are neglected even in the most advanced cosmological simulations of the Schr\"{o}dinger-Poisson equations~\citep{schive2014cosmic,Mocz:2019pyf,Veltmaat2020BaryonDrivenGrowth}: namely, relativistic corrections. A full relativistic cosmological simulation of FDM or ULAs is unfeasible, although Numerial Relativity and other semi-relativsitic models can be used to simulate the relativistic collapse of solitons~\citep{Helfer:2016ljl,Levkov:2016rkk,Michel:2018nzt}. The present work neglects ULA self-interactions, which is a valid approximation in the mass range of interest, where obtaining the correct relic density requires a decay constant $f_a\gtrsim 10^{17}\text{ GeV}$~\citep{Marsh:2015xka}. In this limit solitons undergo collapse to black holes when they exceed the \cite{PhysRev.172.1331} critical mass~\citep{Helfer:2016ljl}. Solitons formed in DM halos by gravitational hierarchical structure formation and obeying the core-halo mass relation~\citep{Schive:2014hza} can be shown to remain always below this critical mass for even the most massive observed halos in the Universe, and thus strong gravity effects can safely be neglected.

A relativistic correction to the Schr\"{o}dinger-Poisson equations that might impact cosmological structure formation is the gravitation of FDM gradient energy, which appears as a correction to the right hand side of the Poisson equation. This correction, and other relativistic corrections, have been studied by \cite{Salehian:2021khb}. The leading relativistic corrections are found to lead to \emph{smaller} solitons than the non-relativistic Schr\"{o}dinger-Poisson solutions considered in the previous section. Thus we can conclude that our approximation to neglect solitons in the halo model based on the non-relativistic model is a conservative one.

%%%%%%%%%%%%%%%%%%%%%%%%%%%%%%%%%%%%%%%%%%%%%%%%%%%%%%%%%%%%%%%%%%%%%%%%%%%%%%%%%
%%%%%%%%%%%%%%%%%%%%%%%%%%%%%%%%%%%%%%%%%%%%%%%%%%%%%%%%%%%%%%%%%%%%%%%%%%%%%%%%%
%
%

%%%%%%%%%%%%%%%%%%%%%%%%%%%%%%%%%%%%%%%%%%%%%%%%%%%%%%
\section{Massive Neutrinos and Other Sources of Power Suppression}\label{appendix:massive_nu}

\begin{figure}
\centering
\includegraphics[width=.48\textwidth]{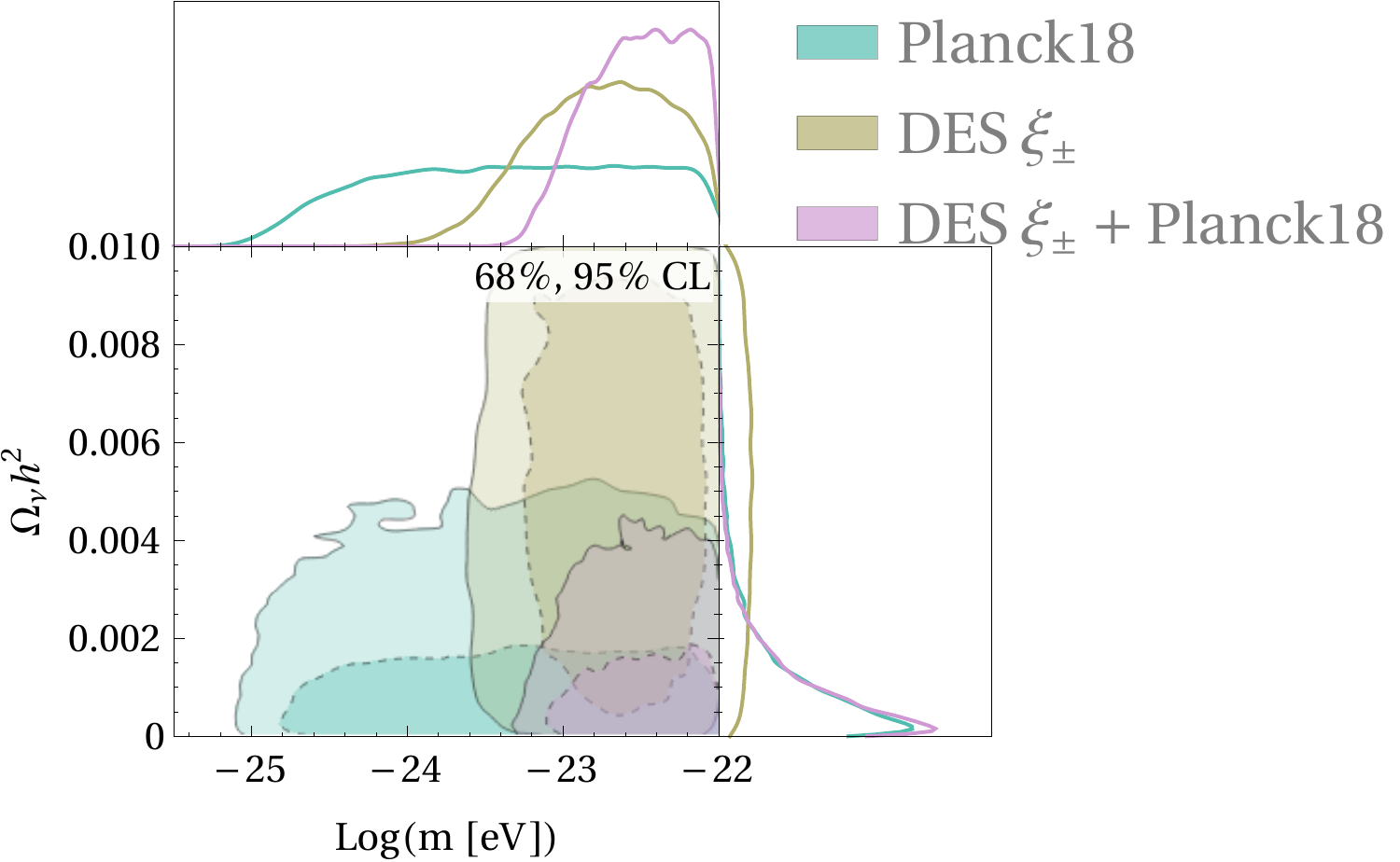}
\caption{Lack of degeneracy between massive neutrinos and ULA mass in our analysis.}
\label{fig:cut_nu_compare}
\end{figure}

Massive neutrinos, due to free streaming, also suppress $P(k)$, leading to step-like features. For observables only sensitive to non-relativistic structure formation, this leads to a degeneracy between the effects of massive neutrinos and ULAs~\citep{Amendola:2005ad,Marsh:2010wq}. For neutrinos in the Standard Model of particle physics for the known cross section, neutrinos freeze-out while they are non-relativistic, and the density and mass are not independent parameters, $\Omega_\nu h^2\propto m_\nu$. Imposing a rough upper limit on the neutrino mass of 1 eV, the power suppression by ULAs only becomes degenerate with neutrinos for $m< 10^{-28}\text{ eV}$, corresponding to field oscillations beginning in the matter-dominated era \citep[``DE-like ULAs'',][]{Hlozek:2014lca}. The degeneracy in $P(k)$ only occurs when ULA fraction is chosen to mimic the $P(k)$ step amplitude of the neutrinos, which demands $\Omega_a\ll 0.12$; alternatively the effective number of neutrinos can be varied along with the ULA density to increase the neutrino density parameter while holding $m_\nu$ fixed~\citep{Marsh:2011bf,Hlozek:2016lzm}. 

Once relativistic observables, in particular the CMB anisotropies, are considered, the degeneracy between massive neutrinos and ULAs is broken. This is due to the different behaviour of ULAs and neutrinos in the relativistic regime, where neutrinos have equation of state $w=1/3$ (hot relativistic particles) and ULAs have $w=-1$ (slowly rolling scalar field), and their consequent differing effects on the expansion rate of the Universe, which leads to differing effects in both the Sachs-Wolfe and Silk damping regions of the CMB power spectrum~\citep{Hlozek:2014lca,Hlozek:2016lzm}.

In the present analysis, we demand that ULAs are all of the DM, which is strongly disfavoured by the data for low masses $m<10^{-28}\text{ eV}$ where the $P(k)$ degeneracy with neutrinos opens up. This is demonstrated in in Fig.~\ref{fig:cut_nu_compare}, where we show the joint posterior on $\Omega_\nu h^2$ and $\log_{10}(m/\text{eV})$. DES data alone does not provide a strong constraint on the neutrino mass~\citep[DES-Y1,][]{Abbott:2017wau}, but demands ULAs have $m\gtrsim 10^{-23}\text{ eV}$, which cannot be degenerate in $P(k)$ for neutrinos with $m_\nu\lesssim 1\text{ eV}$. Including CMB data gives a strong upper bound on the sum of neutrino masses~\citep{Planck:2018vyg} and on the ULA mass when all the DM is ULAs. In the allowed parameter space for the CMB only analysis, ULAs are too heavy to be degenerate with neutrinos. This state of affairs can be understood intuitively: the CMB demands that neutrinos are hot, and become non-relativistic at late times. On the other hand the CMB demands that DM be pressureless at matter radiation equality. Consequently the neutrino free-streaming scale and the ULA Jeans scale are separated by many orders of magnitude and the effects on the matter power spectrum are not degenerate.

\end{document}